\documentclass[10pt]{iopart}
\usepackage{iopams}

\usepackage{gensymb,bbm,wasysym}

\usepackage{graphicx} % Include figure files
\usepackage{bm}       % bold math
\usepackage{tensind}  % for tensor notation
\usepackage{color}
\usepackage{url}
\usepackage{subfigure}

\newcommand{\diff}{\mathrm{d}}

\usepackage{xcolor}

\newcommand{\obs}{\mathrm{obs}}
\newcommand{\uright}{\mathrm{ur}}
\newcommand{\lleft}{\mathrm{ll}}
\newcommand{\sph}{\mathrm{sph}}
\newcommand{\lin}{\mathrm{line}}
\newcommand{\meas}{\mathrm{m}}
\newcommand{\app}{\mathrm{ap}}
\newcommand{\cen}{\mathrm{c}}
\newcommand{\con}{\mathrm{c}}

\newcounter{exampleCounter}
\newenvironment{srExercise}{\stepcounter{exampleCounter}\vspace*{0.1cm} \begin{minipage}[t]{4.5in}\textbf{Exercise \arabic{exampleCounter}:}\it}{\end{minipage}\vspace*{0.1cm}}
\newenvironment{srResult}{\vspace*{0.1cm} \begin{minipage}[t]{4.5in}\textbf{Result:}}{\end{minipage}\vspace*{0.1cm}}
\newenvironment{srConf}{\vspace*{0.1cm} \begin{minipage}[t]{4.5in}\textbf{Configure file: }\color{blue}}{\color{black}\end{minipage}\vspace*{0.1cm}}

\begin{document}
\tensordelimiter{?}

\title{Visual appearance of wireframe objects in special relativity}

\author{Thomas M{\"u}ller}
\address{
  Visualisierungsinstitut der Universit\"at Stuttgart (VISUS)\\
  Allmandring 19, 70569 Stuttgart, Germany
}
\ead{Thomas.Mueller@vis.uni-stuttgart.de}

\author{Sebastian Boblest}
\address{
  Visualisierungsinstitut der Universit\"at Stuttgart (VISUS)\\
  Allmandring 19, 70569 Stuttgart, Germany
}
\ead{Sebastian.Boblest@vis.uni-stuttgart.de}

% -----------------------------------------------------------------
%                            Abstract
% -----------------------------------------------------------------
\begin{abstract}
The visual appearance of a moving object in special relativity can be constructed in a straightforward manner when representing the surface of the object, or at least a wire frame model of it, as a point cloud. 
The apparent position of each individual point is then found by intersecting its worldline with the observer's backward light cone.
In this paper, we present a complete derivation of the apparent position of a point and some more complex geometric objects for general parameter settings (configurations).
We implemented our results in \emph{python} and \emph{asymptote} and used these tools to generate scripts that create the figures in this paper.
These scripts are directly applicable in an undergraduate course to special relativity and can also serve as the basis for student projects with the aim to study more complex sceneries.
\end{abstract}

% -----------------------------------------------------------------
%                            PACS
%
%   http://publish.aps.org/PACS
% -----------------------------------------------------------------

%% 03.30.+p   Special relativity
%% 95.75.Pq   Mathematical procedures and computer techniques

%\submitto{\EJP}

% -----------------------------------------------------------------
%                            Introduction
% -----------------------------------------------------------------
\section{Introduction}\label{sec:intro}
The strange visual appearance of objects moving with a velocity close to the speed of light relative to an observer is one of the puzzling predictions of Einstein's special theory of relativity~\cite{Einstein:1905:EBK} and was studied already by Lampa~\cite{Lampa:1924} in 1924. 
Unfortunately, Lampa's discussion of the apparent shape of a moving rod was not recognized for a long time and even the famous physicist Gamow gave an incorrect conclusion about the visual appearance of a moving wheel in his book ``Mr. Tompkins in Wonderland'' (edition 1940)~\cite{Gamow:1940}.
In 1959, Terrell~\cite{Terrell:1959:InvLC} pointed out that the Lorentz contraction is not visible to an observer, a direction that was similarly pursued by Weinstein~\cite{Weinstein:1960:Obs}, while Penrose~\cite{Penrose:1959:Sphere} proved that a relativistically moving sphere always has a circular outline, a problem that was again considered by Boas~\cite{ Boas:1961:LORS}.         % large object, right circular cone
However, the visual appearance of relativistically moving objects is one of the consequences of special relativity, where an intuitive understanding is hard to reach by performing calculations alone.
The field of relativistic visualization bridges this gap between mathematical results and human imagination.

The rapid increase in computer power and the emergence of very powerful graphics hardware made the development of several sophisticated techniques possible that are capable of generating high quality imagery of special and general relativistic scenarios.

The most natural of these methods is \emph{relativistic ray tracing}, where the physical propagation of light is being reversed and the finite speed of light is taken into account, see for example Hsiung and Dunn~\cite{Hsiung:1989}, Weiskopf~\cite{Weiskopf:2001:Diss}, or M{\"u}ller~\cite{Mueller:2014:GVS}, amongst others.
As this rendering technique is generally very time consuming even on contemporary computers, it is not used for interactive simulations showing the visual effects of relativity. 
A popular alternative method is to transform the polygonal mesh of an object into the observer's rest frame \cite{assol}.
But, this \emph{polygon rendering} technique leads to image artefacts because only the vertices are transformed and the connecting edges are still straight lines.
Ray tracing and polygon rendering can be combined to circumvent the respective disadvantages, however. 
For that, it is necessary to restrict oneself to triangular meshes and make use of the high parallelism of graphics processing units (GPUs) and the free programmability of the graphics pipeline. 
Details of this {\it local ray tracing} technique are described in M{\"u}ller et al.~\cite{Mueller:2010:LRT}.
A recent survey of visualization methods for special relativity was given by Weiskopf~\cite{weiskopf:DFU:2010:2711}.
The reader interested in a comprehensive overview is referred to that paper. 

Among the first visualization techniques employed in special relativity is to consider the apparent shape of wireframe models as such images can be generated also by hand.
Various authors published work that uses this technique. Scott and Viner~\cite{Scott:1965:GA} considered the appearance of plane grids and rectangular boxes,
Scott and van Driel~\cite{Scott:1970:GA} studied, among other things, the look of a sphere passing close to the observer, however without giving a full description of the scenarios they looked at. 
Hickey~\cite{Hickey:1979:Cube} considered the two-dimensional appearance of a relativistically moving cube and  Suffern~\cite{Suffern:1988:Sph} again discussed the outline of a relativistically moving sphere, where he focused on a motion directed towards the observer.
One of the first interactive computer simulations showing the apparent distortion effect at relativistic velocities is {\it Visual Appearance} by Taylor~\cite{Taylor:1989:STS}.
He also uses wireframe objects but does not give any inside in how the visualization is accomplished and his program seems to suffer from polygon rendering artefacts.

In this article, we as well concentrate on wireframes of objects. 
In contrast to earlier work, we not only transform the complete edges in between the vertices according to the Lorentz 
transformations and the finite speed of light so that we can properly visualize how straight lines in general appear bent, but also include depth information to emphasize the apparent shape.
While it is clear that other methods can easily create images of much higher quality, especially by using textures and simple shading techniques, the wireframe method is still very powerful didactically.
On modern computers such visualizations can be created completely interactive and students can create their own sceneries and study the effects of special relativity in these cases.
In this article we give a general derivation of the apparent view of lines and spheres. We allow for a free positioning of these objects in their reference frame, of the observer in his frame, and of the spatial separation and relative speed of the two frames with the only restriction that we assume the axes of the two frames to be aligned.
This allows to construct complex scenes on the one hand and to study how different observers perceive the same scenery on the other hand.

Our results are implemented in \emph{asymptote}~\cite{asymptote} and \emph{python} scripts that we used to create the figures in this paper but which, more importantly, may be used in courses to special relativity or in student projects where other scenes could be constructed and studied. 
Our scripts can be downloaded from \url{http://go.visus.uni-stuttgart.de/srwireframe}. 
With the \emph{python} scripts, some scenes can also be animated.

The structure of this paper is as follows. 
In section~\ref{sec:poincare} we recapitulate the Poincar\'e transformation that is the basis for all further calculations.
In section~\ref{sec:appPos} we give a detailed mathematical derivation of the parametrized equations for the apparent view of a single point, a rod, and a sphere. 
In section~\ref{sec:examples}, we specialize to some descriptive examples and compare our wireframe models with the corresponding rendered images which follow from four-dimensional ray tracing.
\ref{appsec:examples} gives some further examples in forms of exercises.

% -----------------------------------------------------------------
%                    Poincare transformation
% -----------------------------------------------------------------
\section{Poincar\'e Transformation}\label{sec:poincare}
\begin{figure}[ht]
  \centering
  \includegraphics[width = 0.6\textwidth]{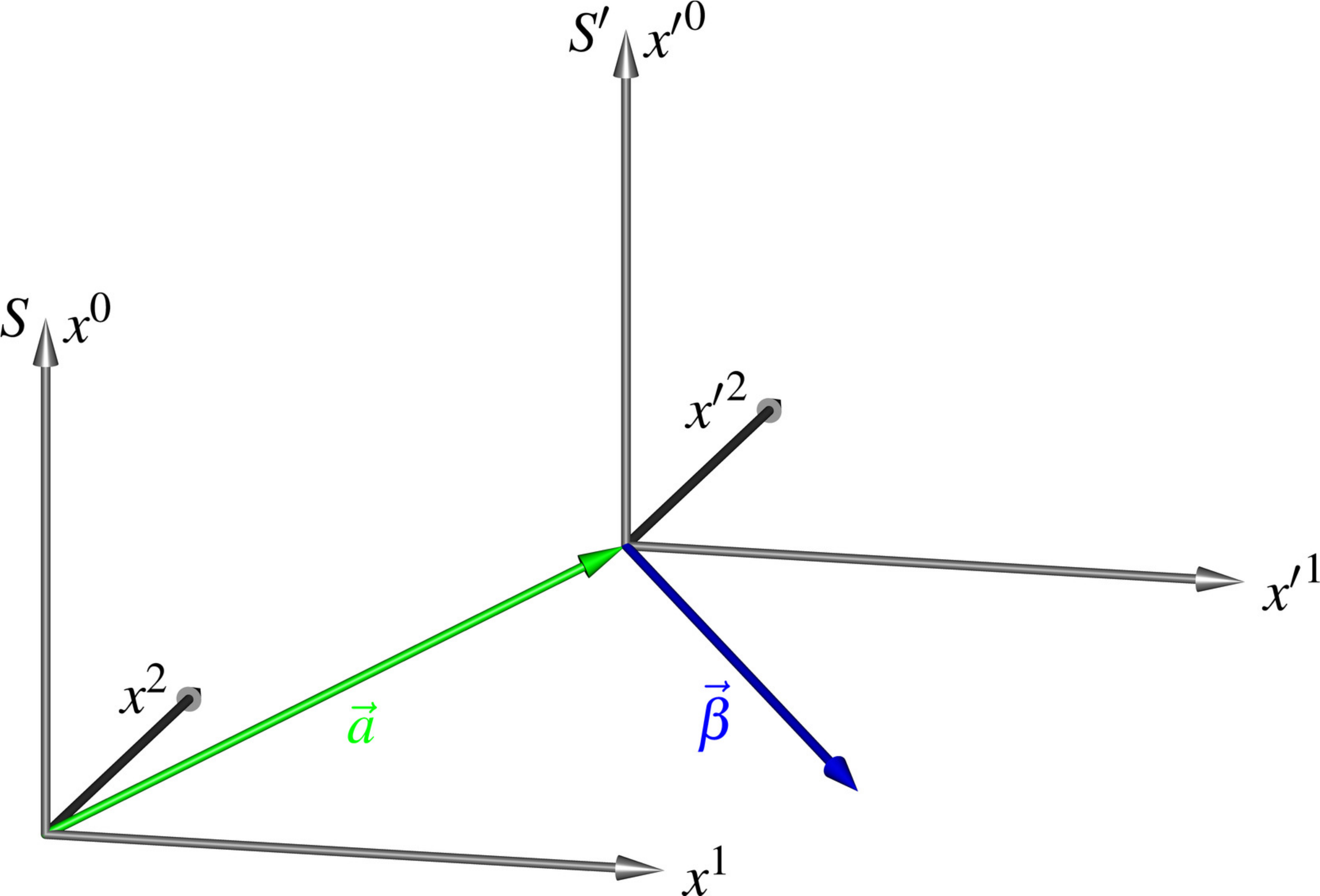}
   \caption{The frame of reference $S'$ is moving with constant velocity $\vec{\beta}$ with respect to $S$. Both systems are synchronized to $x^0=x'^0=0$ when $S'$ is located at $\vec{a}$ with respect to $S$; and their axes are aligned.}
 \label{fig:standard}
\end{figure}

Consider two frames of reference $S$ and $S'$ equipped with their individual coordinate systems $x^{\mu}=(x^0,x^1,x^2,x^3)=(ct,\vec{x})$ and $x'^{\mu}=(x'^0,x'^1,x'^2,x'^3) = (ct',\vec{x}')$, respectively.
The clocks of both frames are synchronized to $x^0=x'^0=0$ when the origin of $S'$ is located at $\vec{a}$ with respect to the origin of $S$, see figure~\ref{fig:standard}. 
The coordinate axis of both frames are aligned to each other and $S'$ moves with constant velocity $\vec{\beta}$ with respect to $S$.
We will refer to this setup as being the \emph{standard configuration} in special relativity without rotations.
Our observer will be at rest in the system $S$, while the system $S'$ is the rest frame for our sceneries.

The Poincar{\'e} transformation between both frames is defined by
\begin{equation}
 \label{eq:poincareSp2S}
 x^{\mu} = ?\Lambda^{\mu}_{\nu}?x'^{\nu} + a^{\mu},
\end{equation}
and the Lorentz matrix $?\Lambda^{\mu}_{\nu}?$ is given by
\begin{equation}
  ?\Lambda^0_0? = \gamma,\quad ?\Lambda^0_i?=\gamma\beta_i,\quad ?\Lambda^i_0?=\gamma\beta^i,\quad  ?\Lambda^i_j? = \delta^i_j+\frac{\gamma^2}{\gamma+1}\beta^i\beta_j,
  \label{eq:Lorentz}
\end{equation}
where $\beta_i\beta^i=\beta^i\beta^i=\vec{\beta}\cdot\vec{\beta}<1$ and $\gamma=1/\sqrt{1-\beta_i\beta^i}$, see e.g. Misner et al.~\cite{mtw}, and $\delta^i_j$ is the Kronecker-$\delta$.
We also use Einstein's sum convention to sum over indices that appear twice in the 
same term.
The displacement four-vector reads $a^{\mu}=(0,\vec{a})$.
Here and in the following, Greek indices run from $0$ to $3$, where the $0$-th coordinate represents time, and Latin indices go from $1$ to $3$. 

The inverse of the Lorentz matrix $\bar\Lambda=\Lambda^{-1}$ reads
\begin{equation}
  ?{\bar\Lambda}^0_0? = \gamma,\quad ?{\bar\Lambda}^0_i?=-\gamma\beta_i,\quad ?{\bar\Lambda}^i_0?=-\gamma\beta^i,\quad  ?{\bar\Lambda}^i_j? = \delta^i_j+\frac{\gamma^2}{\gamma+1}\beta^i\beta_j,
    \label{eq:LorentzInv}
\end{equation}
which differs from the initial Lorentz matrix only by the sign of the velocity $\vec{\beta}$.
The corresponding Poincar\'e transformation reads $x'^{\mu}=?{\bar\Lambda}^{\mu}_{\nu}?(x^{\nu}-a^{\nu})$.

% -----------------------------------------------------------------
%                    Apparent position
% -----------------------------------------------------------------
\section{Apparent view of an object}\label{sec:appPos}
The finite speed of light is responsible for the fact that we do not see a moving object where it actually is, but where it was when it sent the light that we now observe. 

% --------------------------------------------
%   Apparent position of a point
% --------------------------------------------
\subsection{Apparent position of a point}\label{sec:appPoint}
In the simplest case, the object is just a point $P$ and its apparent position can be determined by intersecting the point's worldline $\left(x_p^0,\vec{x}_p=\vec{x}_p(x_p^0)\right)$ with the backward light cone of the observer, who is static with respect to $S$,
\begin{equation}
  0 = -\left(x^0-x^0_{\obs}\right)^2 + \sum\limits_{i=1}^3\left(x^i-x_{\obs}^i\right)^2.
  \label{eq:backwardLightcone}
\end{equation}
(For the rest of this paper, we drop the index $p$ of the point.)

If $P$ is at rest with respect to the moving frame $S'$, $\vec{x}'=\mbox{const}=\vec{x}'_p$, we have to transform its worldline into $S$ by means of the Poincar\'e transformation (\ref{eq:poincareSp2S}). 
Then, equation (\ref{eq:backwardLightcone}) yields
\begin{eqnarray}
\fl 0 &= -\left(?\Lambda^0_{\nu}?x'^{\nu}-x_{\obs}^0\right)^2 + \delta_{ij}\left(?\Lambda^i_{\nu}?x'^{\nu}+a^i-x_{\obs}^i\right)\left(?\Lambda^j_{\mu}?x'^{\mu}+a^j-x_{\obs}^j\right)\\
\nonumber\fl  &= -\left(?\Lambda^0_0?x'^0+?\Lambda^0_n?x'^n-x_{\obs}^0\right)^2\\
\fl &\quad + \delta_{ij}\left(?\Lambda^i_0?x'^0+?\Lambda^i_n?x'^n+a^i-x_{\obs}^i\right)\left(?\Lambda^j_0?x'^0+?\Lambda^j_m?x'^m+a^j-x_{\obs}^j\right).
  \label{eq:blc2}
\end{eqnarray}
Here, the only unknown is $x'^0$ which is the time when light must be emitted by the point in order to reach the observer at time $x^0_{\obs}$. 
Solving the quadratic equation (\ref{eq:blc2}) for $x'^0$ and using the abbreviations
\begin{eqnarray}
  \rho &= \gamma\left(\vec{\beta}\cdot\vec{x}'\right) - x^0_{\obs}, \quad
  \vec{\eta} = \vec{x}'+\frac{\gamma^2}{\gamma+1}\left(\vec{\beta}\cdot\vec{x}'\right)\vec{\beta} + \vec{a} -\vec{x}_{\obs},  \label{eq:rho_eta} \\
  \omega_0^2 &= \gamma^2(\vec{\beta}\cdot\vec{\eta}-\rho)^2-\rho^2+\vec{\eta}\cdot\vec{\eta}, \label{eq:omega0}
\end{eqnarray}
yields
\begin{eqnarray}
  \label{eq:is1a} x'^0 &= \gamma\left(\vec{\beta}\cdot\vec{\eta}-\rho\right)- \omega_0, \\
  \label{eq:is1b} \vec{x} &= \vec{x}' +\left(\gamma x'^0 + \frac{\gamma^2}{\gamma+1}\vec{\beta}\cdot\vec{x}'\right)\vec{\beta} + \vec{a} = \gamma 
x'^0\vec{\beta}+\vec{\eta}+\vec{x}_{\obs}.
  \label{eq:appSinglePoint}
\end{eqnarray}
Note that in these expressions the scalar product as usual is an abbreviation for the sum over all products of the vector components, like for example,  $\vec{\beta}\cdot\vec{\eta}=\beta^1\eta^1+\beta^2\eta^2+\beta^3\eta^3$.
However, it must not be interpreted with respect to either one of the reference frames $S$ or $S'$, respectively.
Hence, it has to be taken by care how to interpret the situation when the scalar product vanishes, $\vec{\beta}\cdot\vec{\eta}=0$.
In general, it cannot be interpreted as both vectors being ``perpendicular'', because some of them are a mixture of vectors measured with respect to $S$ or $S'$.

Equations (\ref{eq:is1a}) and (\ref{eq:is1b}) simplify considerably if the point $P$ and the observer are in the origin 
of their respective reference frames, i.e. $\vec{x}' = \vec{0} = \vec{x}_{\obs} $, and the displacement vector $\vec{a}=\vec{0}$.
Then, 
\begin{equation}
  x'^0 = \gamma(1\pm\beta) x_{\obs}^0 = \sqrt{\frac{1\pm\beta}{1\mp\beta}}x_{\obs}^0=:D_{\beta}x_{\obs}^0 \quad\mbox{and}\quad \vec{x} = \gamma x'^0\vec{\beta}.
\end{equation}
While $x_{\obs}^0<0$, the point approaches the observer and we have to use the upper signs in the square root factor $D_{\beta}$.
After $P$ has passed the observer, we have to use the lower signs, respectively.
$D_{\beta}$ is also called \emph{Doppler factor} and is responsible for a blue- or red-shift if the spectrum of the light would be taken into consideration.

\begin{figure}[ht]
  \centering
  \includegraphics[width = 0.7\textwidth]{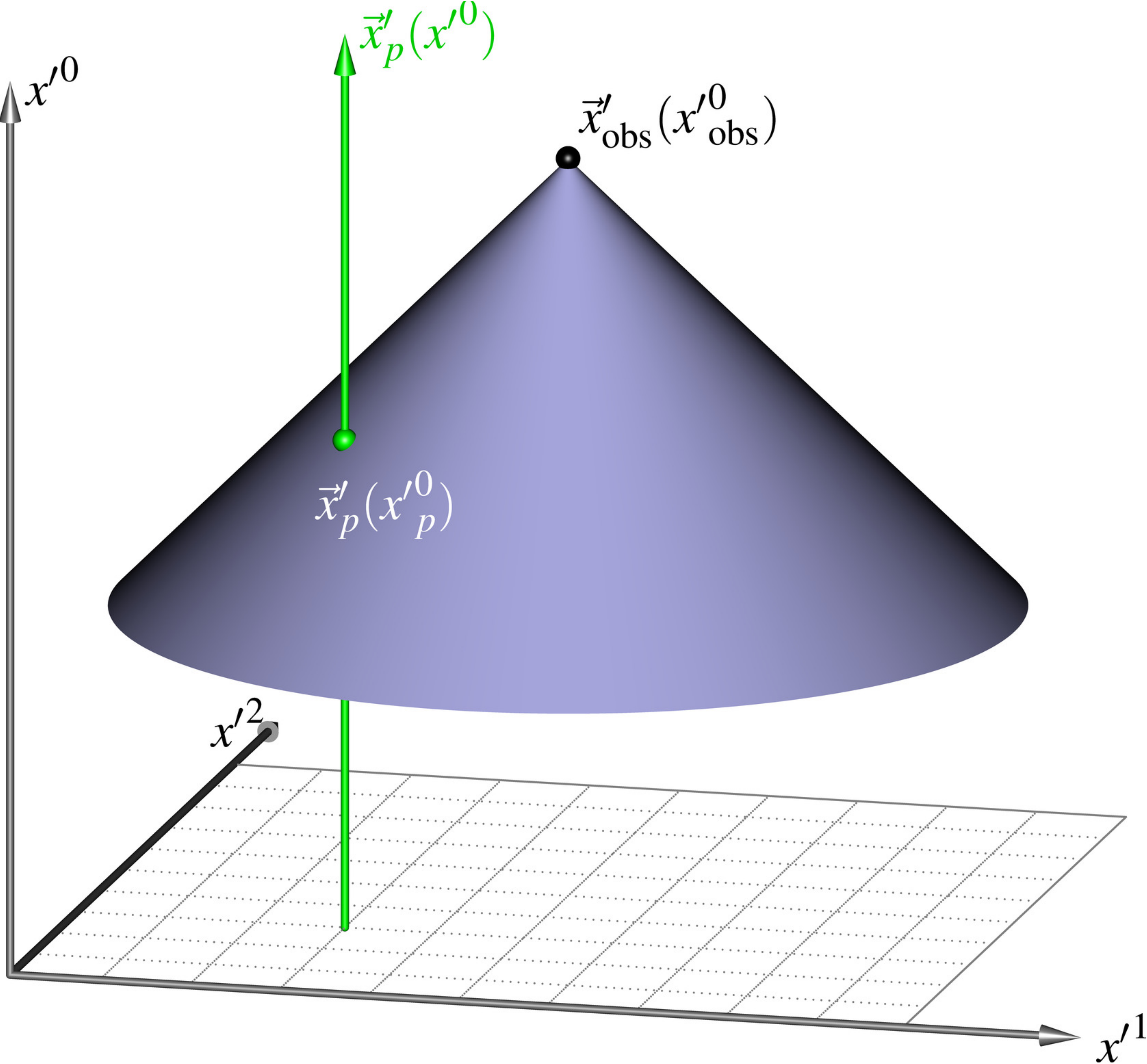}
  \caption{The intersection between the worldline of the point $P$ and the backward light cone of the observer determines the event $\vec{x}'_p({x'}_p^0)$, where $P$ has to emit light that is seen by the observer at time ${x'}_{\obs}^0$.
  }
  \label{fig:lightcone}
\end{figure}

We could also accomplish the light cone intersection within the frame $S'$, where the observer's current position at their observation time $x_{\obs}^0$ follows from the inverse Poincar\'e transformation, $x'^{\mu}_{\obs} = ?{({\Lambda^{-1}})}^{\mu}_{\nu}?\left(x^{\nu}_{\obs}-a^{\nu}\right)$, see figure~\ref{fig:lightcone}. 

Then, the intersection of the light cone 
\begin{equation}
 0 = -\left(x'^0-x'^0_{\obs}\right)^2+\sum\limits_{i=1}^3\left(x'^i-x'^i_{\obs}\right)^2
 \label{eq:lightconeSs}
\end{equation}
with the static point $P$ immediately yields
\begin{equation}
 x'^0 = x'^0_{\obs} - \Delta(\vec{x}',\vec{x}'_{\obs}), \quad \Delta(\vec{x}',\vec{x}'_{\obs}) := \sqrt{\sum\limits_{i=1}^3\left(x'^i-x'^i_{\obs}\right)^2}.
 \label{eq:intersecTimeS}
\end{equation}
The apparent position $\vec{x}$ follows from the back transformation by means of equation~(\ref{eq:poincareSp2S}). 
This second approach appears to be more straight, but it needs two Poincar\'e transformations. 

With the above transformations at hand, we could determine the virtual shape of any relativistically moving object by means of representing its surface by a cloud of points. 
The resulting apparent positions make up the \emph{photo-object}, which is the set of all points where light is emitted from the object's surface that reaches the observer at the same time.
This photo-object is what the observer or its camera will \emph{see}.
However, the observer's perception might differ from what he sees and depends also on the texture of an object; we will come to this point later in Sec.~\ref{subsec:expAppViewBall}.

% --------------------------------------------
%    Apparent view of a line/rod       
% --------------------------------------------
\subsection{Apparent view of a line/rod}\label{subsec:appPosLine}
Instead of a single point, we now consider a straight line segment $\vec{x}'+s'\vec{\sigma}'$, $s'\in[s'_1,s'_2]$, that is defined by a specific reference point $\vec{x}'$ 
and a direction $\vec{\sigma}'$ with $\|\vec{\sigma}'\|=1$ as measured in $S'$. 
Replacing $\vec{x}'$ in equation~(\ref{eq:appSinglePoint}) by $\vec{x}'+s'\vec{\sigma}'$ yields
\begin{eqnarray}
 \label{eq:appLineX}
 \vec{x}\left(s',\vec{\sigma}'\right) &= \gamma x'^0\left(s',\vec{\sigma}'\right)\vec{\beta}+\vec{\eta}+\vec{x}_{\obs}+s'\vec{\sigma}' + \frac{\gamma^2s'}{\gamma+1}\left(\vec{\beta}\cdot\vec{\sigma}'\right)\vec{\beta},\\
 \label{eq:appLineT} x'^0\left(s',\vec{\sigma}'\right) &= \gamma\left(\vec{\beta}\cdot\vec{\eta}-\rho\right)-\omega_{\lin}(s'),
\end{eqnarray}
where $\rho$ and $\vec{\eta}$ are the same abbreviations as in (\ref{eq:rho_eta}), and 
\begin{eqnarray}
  \label{eq:omega}\omega_{\lin}(s')^2 &= \omega_0^2 +2s'\upsilon+s'^2,\\
  \label{eq:mu}\upsilon &= \vec{\mu}\cdot\vec{\sigma}',\quad\mbox{with}\quad\vec{\mu} = \vec{\eta} + \vec{\beta}\left(-\gamma\rho+\frac{\gamma^2}{\gamma+1}\vec{\beta}\cdot\vec{\eta}\right),
\end{eqnarray}
with $\omega_0$ from equation (\ref{eq:omega0}).
Each point of a line has to emit light at a different time such that it is being received by the observer at their observation time. Hence, the apparent shape of the line will not be straight, in general. Details can be determined using the \emph{Frenet-Serret frame} along the line segment defined by the tangent $\vec{e}_1(s')$, the main normal $\vec{e}_2(s')$, and the binormal $\vec{e}_3(s')=\vec{e}_1(s')\times\vec{e}_2(s')$, where
\begin{equation}
  \vec{e}_1(s') = \frac{\diff\vec{x}(s')/\diff{s'}}{\|\diff\vec{x}(s')/\diff{s'}\|}
\end{equation}
with derivative
\begin{eqnarray}
  \frac{\diff\vec{x}(s')}{\diff{s'}} = -\gamma\frac{s'+\upsilon}{\omega_{\lin}(s')}\vec{\beta}+\vec{\sigma}'+\frac{\gamma^2}{\gamma+1}\left(\vec{\beta}\cdot\vec{\sigma}'\right)\vec{\beta}
\end{eqnarray}
and corresponding norm
\begin{eqnarray}
  \fl\bigg\|\frac{\diff\vec{x}(s')}{\diff{s'}}\bigg\|^2 &= \gamma^2\frac{(s'+\upsilon)^2}{\omega_{\lin}(s')^2}\beta^2-2\gamma^2\frac{s'+\upsilon}{\omega_{\lin}(s')}\left(\vec{\beta}\cdot\vec{\sigma}'\right)+1+\gamma^2\left(\vec{\beta}\cdot\vec{\sigma}'\right)^2 :=f(s')^2.
\end{eqnarray}
Note that $\vec{e}_1,\vec{e}_2,\vec{e}_3$ are parametrized by $s'$ but are given with respect to the frame $S$.

As the Frenet-Serret frame is only valid for a curve parametrized by its arc length, the main normal cannot be determined directly from the second derivative of equation~(\ref{eq:appLineX}), but has to be calculated from the derivative of the tangent $\vec{e}_1(s')$. Thus, $\vec{e}_2(s')=(\diff\vec{e}_1/\diff{s'})/\|\diff\vec{e}_1/\diff{s'}\|$ with
\begin{eqnarray}
  \fl\nonumber\frac{\diff\vec{e}_1}{\diff{s'}} &= -\frac{\omega_{\lin}(s')^2-(s'+\upsilon)^2}{f(s')^3\omega_{\lin}(s')^3}\bigg\{\vec{\beta}\left[-\gamma^2\frac{s'+\upsilon}{\omega_{\lin}(s')}(\vec{\beta}\cdot\vec{\sigma}')+\gamma+\frac{\gamma^3}{\gamma+1}(\vec{\beta}\cdot\vec{\sigma}')^2\right]\\
  \fl &\quad + \vec{\sigma}'\gamma^2\left[\beta^2\frac{s'+\upsilon}{\omega_{\lin}(s')}-(\vec{\beta}\cdot\vec{\sigma}')\right]\bigg\}.
   \label{eq:de1ds}
\end{eqnarray}
The absolute value of (\ref{eq:de1ds}) not only yields the normalization factor for the main normal but it also yields the curvature of the curve, $\kappa(s')=\|\diff\vec{e}_1/\diff{s'}\|$, which is given by
\begin{eqnarray}
  \label{eq:curvature}
  \fl\kappa(s')^2 &= \left[\frac{\omega_{\lin}(s')^2-(s'+\upsilon)^2}{f(s')^3\omega_{\lin}(s')^3}\right]^2\gamma^2\bigg\{\left(\frac{s'+\upsilon}{\omega_{\lin}(s')}\right)^2\gamma^2\beta^2\left[\beta^2-(\vec{\beta}\cdot\vec{\sigma}')^2\right]\\
  \fl\nonumber &-2\gamma^2(\vec{\beta}\cdot\vec{\sigma}')\frac{s'+\upsilon}{\omega_{\lin}(s')}\left[\beta^2-(\vec{\beta}\cdot\vec{\sigma}')^2\right]+\left[\beta^2+(\gamma^2-2)(\vec{\beta}\cdot\vec{\sigma}')^2-\gamma^2(\vec{\beta}\cdot\vec{\sigma}')^4\right]\bigg\}.  
\end{eqnarray}
The exact form of the binormal $\vec{e}_3(s')$ is of no interest here. 

In the special case $\vec{\sigma}'\parallel\vec{\beta}$, or $\vec{\beta}=\pm\beta\vec{\sigma}'$, respectively, the straight line points in the direction of motion, and the curvature in this parallel case $\kappa^{\parallel}(s')\equiv 0$ for all $s'\in\mathbb{R}$. 
Hence, such lines only change their apparent length but do not appear to be bent, see also the example in Sec.~\ref{subsec:expAppViewLineAlong}.

If $\vec{\beta}\cdot\vec{\sigma}'=0$, the curvature, equation~(\ref{eq:curvature}), simplifies considerably.
Then,
\begin{equation}
  \kappa^{\perp}(s') = \frac{\omega_{\lin}(s')^2-(s'+\upsilon)^2}{f(s')^2\omega_{\lin}(s')^3}\gamma\beta
  \label{eq:curvPerp}
\end{equation}
with maximum given at $s'_{\max}=-\upsilon=-\vec{\eta}\cdot\vec{\sigma}'$. 
Thus, $\kappa^{\perp}_{\max}=\gamma\beta/\omega_{\lin}(-\upsilon)$. 
As to be expected, the tangent and the main normal read $\vec{e}_1^{\perp}(-\upsilon)=\vec{\sigma}'$ and $\vec{e}_2^{\perp}(-\upsilon)=-\vec{\beta}/\beta$, respectively. 
The apparent line, equation~(\ref{eq:appLineX}), reduces to
\begin{equation}
  \vec{x}^{\perp}(s',\vec{\sigma}')=\vec{p}-\gamma\omega_{\lin}(s')\vec{\beta} + s'\vec{\sigma}'
\end{equation}
with the reference point $\vec{p}=\gamma^2[x_{\obs}^0+\vec{\beta}\cdot(\vec{a}-\vec{x}_{\obs})]\vec{\beta}+\vec{\eta}+\vec{x}_{\obs}$. To show that $\vec{x}^{\perp}(s',\vec{\sigma}')$ has the form of a hyperbola, the local coordinates $(\zeta,\xi)$ with respect to the coordinate system spanned by $\vec{\beta}$ and $\vec{\sigma}'$ are defined. 
With $\xi=\vec{\sigma}'\cdot\vec{x}^{\perp}=\vec{\sigma}'\cdot\vec{p}+s'=\xi_0+s'$ and $\zeta=\vec{\beta}/\beta\cdot\vec{x}^{\perp}=\vec{\beta}/\beta\cdot\vec{p}-\gamma\beta\omega_{\lin}(s')$, the ansatz $(\zeta-\zeta_0)^2/a^2-(\xi-c)^2/b^2=1$ yields
\begin{equation}
  1 = \frac{(\zeta-\zeta_0)^2}{\gamma^4\beta^4/(\kappa^{\perp})^2}-\frac{(\xi-\xi_0+\upsilon)^2}{\gamma^2\beta^2/(\kappa^{\perp})^2},
\end{equation}
where $\zeta_0=\gamma^2\beta[x_{\obs}^0+\vec{\beta}\cdot(\vec{a}-\vec{x}_{\obs})]+\gamma(\vec{\beta}\cdot\vec{x}')+\vec{a}\cdot\vec{\beta}$ and $\xi_0=(\vec{x}'+\vec{a})\cdot\vec{\sigma}'$.

The osculating circle at the point of maximum curvature has radius $(\kappa^{\perp})^{-1}$ and is centred at $\vec{m}=\vec{x}^{\perp}(-\upsilon,\vec{\sigma})+(\kappa^{\perp})^{-1}\vec{e}_2^{\perp}(-\upsilon)$. 
The corresponding local coordinates read $\zeta_m=\zeta_0-\gamma^2(\kappa^{\perp})^{-1}$ and $\xi_m=\vec{x}_{\obs}\cdot\vec{\sigma}'$.

% --------------------------------------------
%   Apparent view of a sphere
% --------------------------------------------
\subsection{Apparent view of a sphere}\label{sec:appViewSphere}
The surface of a sphere within the reference frame $S'$ can be defined by the central point $\vec{x}'$, the orthonormal basis vectors $\{\vec{\sigma}'_1,\vec{\sigma}'_2,\vec{\sigma}'_3\}$, the radius $r'$, and the spherical coordinates $\vartheta'\in(0,\pi)$ and $\varphi'\in[0,2\pi)$. 
An approach similar to the one for the line, where $\vec{x}'$ is now replaced by $\vec{x}'+r'\sin\vartheta'\cos\varphi'\vec{\sigma}'_1+r'\sin\vartheta'\sin\varphi'\vec{\sigma}'_2+r'\cos\vartheta'\vec{\sigma}'_3=\vec{x}'+\sum_{i=1}^3 s'_i\vec{\sigma}'_i$ in equation~(\ref{eq:appSinglePoint}), yields,
\begin{eqnarray}
  \vec{x} &= \gamma x'^0\vec{\beta}+\vec{\eta}+\vec{x}_{\obs}+\sum\limits_{i=1}^3s'_i\vec{\sigma}'_i+\frac{\gamma^2}{\gamma+1}\left(\sum\limits_{i=1}^3s'_i\vec{\sigma}'_i\cdot\vec{\beta}\right)\vec{\beta}, \label{eq:appSpherePosX}\\
  x'^0 &= \gamma\left(\vec{\beta}\cdot\vec{\eta}-\rho\right) - \omega_{\sph}(s'_1,s'_2,s'_3),\label{eq:appSpherePosT}
\end{eqnarray}
where
\begin{eqnarray}
  \omega_{\sph}(s'_1,s'_2,s'_3)^2 &= \omega_0^2 + 2\vec{\mu}\cdot\sum\limits_{i=1}^3s'_i\vec{\sigma}'_i + r'^2,
\end{eqnarray}
and $\vec{\mu}$ is the same expression as in equation (\ref{eq:mu}). 

Equations~(\ref{eq:appSpherePosX}) and (\ref{eq:appSpherePosT}) simplify considerably if the sphere's center point $\vec{x}'$, the observer position $\vec{x}_{\obs}$, and the system offset $\vec{a}$ vanish identically.
Additionally, the basis vectors $\vec{\sigma}'_1$, $\vec{\sigma}'_2$, and $\vec{\sigma}'_3$ are equal to the standard basis vectors $\vec{e}'_1=(1,0,0)^T$, $\vec{e}'_2=(0,1,0)^T$, and $\vec{e}'_3=(0,0,1)^T$, and the velocity $\vec{\beta}=(\beta,0,0)^T$ has only a non-vanishing component in the $x^1$-direction.
Then,
\begin{eqnarray}
  \omega_{\sph}^2 &=& \gamma^2\beta^2\left(x_{\obs}^0\right)^2 + 2\gamma x_{\obs}^0\beta r'\sin\vartheta'\cos\varphi' + r'^2,\\
  x^1 &=& \gamma \left(\gamma x_{\obs}^0 - \omega_{\sph}\right)\beta + \gamma r'\sin\vartheta'\cos\varphi',\\
  x^2 &=& r'\sin\vartheta'\sin\varphi',\\
  x^3 &=& r'\cos\varphi'.
\end{eqnarray}
As expected, the $x^2$- and $x^3$-components are not influenced, because the sphere only moves along the $x^1$-direction and the other parameters are like in the standard literature.

The silhouette of a sphere always appears to be circular irrespective of the sphere's motion, as shown already by others. 
We give a short sketch in \ref{appsec:silhouette} of how this could be proven.

% -----------------------------------------------------------------
%                    Examples
% -----------------------------------------------------------------
\section{Examples}\label{sec:examples}
In the following, we will present some typical examples.
All of them can be reproduced by the accompanying \emph{asymptote} and \emph{python} scripts.
We also compare the wireframe representations with the corresponding images rendered using the four-dimensional ray tracing code \emph{GeoViS}~\cite{Mueller:2014:GVS}.
The great advantage of the python scripts is the possibility to animate the scenes without delay while ray tracing codes might take several minutes to render an image sequence which has to be concatenated into a film afterwards.
Besides the script names mentioned in the figure captions, we use the common script \texttt{sr\_apparent} that contains the calculation of the apparent positions discussed in the previous sections.
Note that script names without file ending are valid for asymptote as well as python.

% ----------------------------------------------
%   Eye or camera transformation
% ----------------------------------------------
\subsection{Eye or camera transformation}
In section~\ref{sec:appPos}, we deduced the apparent position of a single point, a point on a line, or a point on a sphere.
This apparent position is the position in space where light has to be emitted by the point in order to reach the observer at their observation time.
The next step is to map the apparent position of the point into the eye or the camera of the observer which we both represent by a pinhole camera.
For that, we first transform the apparent position into the camera's standard reference frame by means of the \emph{View} matrix.
Then, the \emph{Projection} matrix emulates the perspective projection of the pinhole camera.
The \emph{View} and \emph{Projection} matrices are defined in \texttt{sr\_camera}, see also \ref{appsec:matrices}.
For further details, we refer the reader to the standard literature of computer graphics like, e.g., Foley~\cite{Foley:1996:CG} or Shirley et al.~\cite{Shirley:2009:FGC}.

% ----------------------------------------------
%   Apparent view of a line/rod along
% ----------------------------------------------
\subsection{Apparent view of a line/rod oriented along its direction of motion}\label{subsec:expAppViewLineAlong}
\begin{figure}[ht]
  \subfigure[Minkowski diagram, $\beta=0.5$]{\label{fig:minkRod}\includegraphics[height=170px]{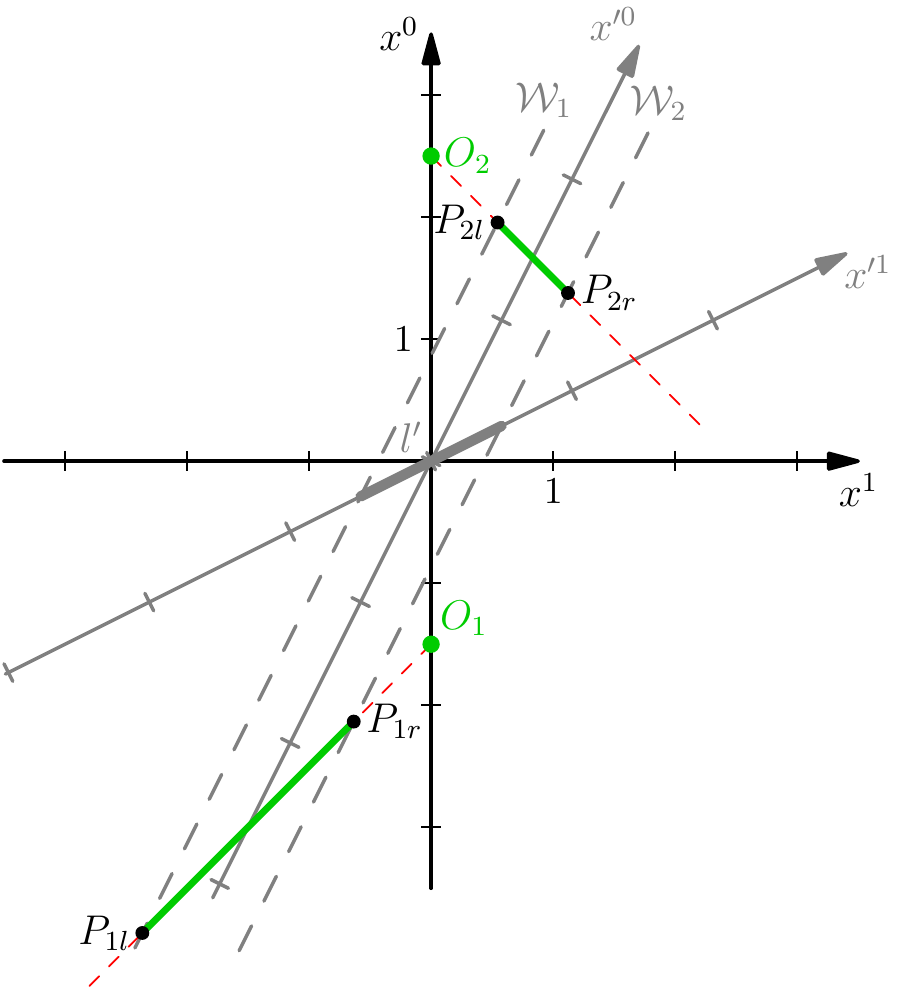}}\hfill
  \subfigure[Length depending on $\beta$]{\label{fig:appLengthBeta}\includegraphics[height=170px]{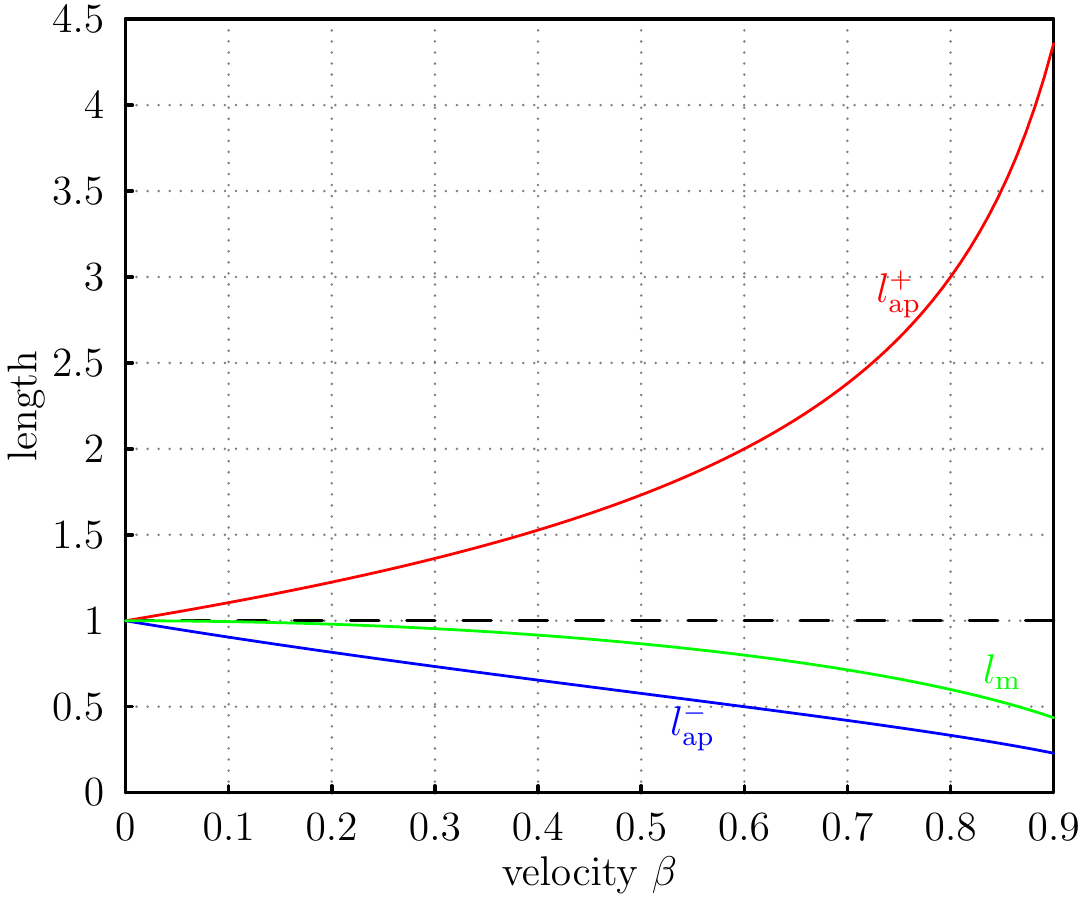}}
  \caption{(a) A rod of length $l'=1$ with respect to $S'$ is oriented along its direction of motion $\vec{\beta}=\beta\vec{\sigma}'$, where $\vec{\sigma}'$ points in the positive $x^1$-direction of $S$. 
  In this Minkowski diagram, the projections of the green lines onto the $x$-axis yield the apparent lengths of the rod.
  (b) Apparent length $l_{\app}^{\pm}$ of the approaching (red) and the receding (blue) rod. 
  The green line represents the measured length $l_{\meas}$.
  }  
\end{figure}

The most fundamental object besides a point is a straight line or rod.
If the rod's orientation is alongside its direction of motion, then equations (\ref{eq:appLineX}) and (\ref{eq:appLineT}) can be simplified.
Thus, with $\vec{\beta}=\beta\vec{\sigma}'$, $\vec{a}=\vec{0}$, $\vec{x}'=\vec{0}$ and $s'\in[-l'/2,l'/2]$, we obtain
\begin{eqnarray}
  \fl\vec{x}(s',\vec{\sigma}') &=& \gamma\left[x'^0(s',\vec{\sigma}')\beta+s'\right]\vec{\sigma}',\\
  \fl x'^0(s',\vec{\sigma}') &=& \gamma\left(-\beta\vec{\sigma}'\cdot\vec{x}_{\obs}+x_{\obs}^0\right)-\omega_{\lin}(s'),\\
  \fl\omega_{\lin}(s')^2 &=& \gamma^2\left(-\beta\vec{\sigma}'\cdot\vec{x}_{\obs}+x_{\obs}^0\right)^2-\left(x_{\obs}^0\right)^2+\|\vec{x}_{\obs}\|^2+2s'\upsilon+s'^2,\\
  \fl \upsilon &=& \gamma\left(\beta x_{\obs}^0-\vec{\sigma}\cdot\vec{x}_{\obs}\right).
\end{eqnarray}
If additionally $\vec{x}_{\obs}=\vec{0}$, the direction $\vec{\sigma}'$ is insignificant, and the rod can only move towards or away from the observer.
Then, the apparent length $l_{\app}^2=\|\vec{x}(l'/2,\vec{\sigma}')-\vec{x}(-l'/2,\vec{\sigma}')\|^2$ of the rod is given by
\begin{equation}
  l_{\app}^{\pm} = \gamma l'\left(1\pm\beta\right),
\end{equation}
where the upper (lower) sign represents the approaching (receding) rod.
This is also true for the slightly more general case $\vec{x}_{\obs}=\xi\vec{\sigma}$.
The Minkowski diagram, figure~\ref{fig:minkRod}, depicts this situation.
At observation event $O_1$, the apparent length of the approaching rod is determined by the $x^1$-coordinates of the events $P_{1l}$ and $P_{1r}$. Thus, for $\beta=0.5$, we obtain $l_{\app}^{+}\approx 1.732~l'$.
At $O_2$, the rod recedes from the observer and has an apparent length $l_{\app}^{-}\approx 0.577~l'$.
% (Script: \texttt{minkRodMoving.asy})
The length $l_{\meas}$, measured by two observers who are in synchronicity with respect to $S$, follows from the Lorentz-Fitzgerald contraction equation: $l_{\meas}=l'/\gamma$.
Thus, a rod which approaches the observer always appears longer than it actually is.
A receding rod, however, appears to be even shorter than its measured length with respect to $S$.
Figure~\ref{fig:appLengthBeta} shows the apparent lengths $l_{\app}^{\pm}$ and the measured length $l_{\meas}$ both as functions of the velocity $\beta$.

Strictly speaking, if a line (rod) is oriented alongside its direction of motion, only a point (the tip) is visible.
Even if the line or rod is slightly off-axis, the perspective projection has to be taken into account which prevents the observer from seeing the calculated apparent lengths.

% ----------------------------------------------
%   Apparent view of a die
% ----------------------------------------------
\subsection{Apparent view of a die}\label{subsec:expAppViewDie}
\begin{figure}[ht]
 \subfigure[Rendered, $\vec{x}_{\obs} = (14.422,0,2)^T$]{\includegraphics[width=0.497\textwidth]{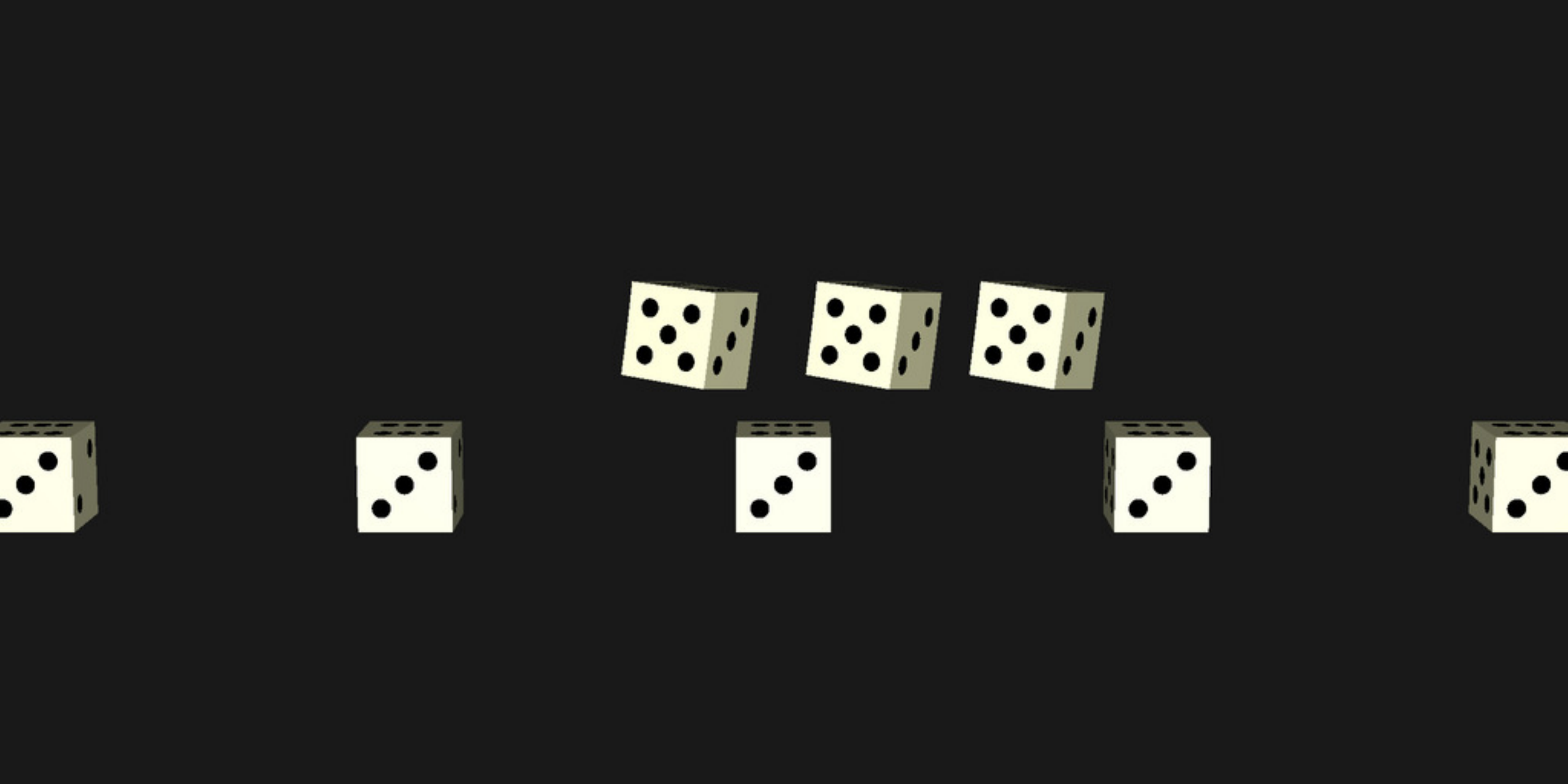}}\hfill
 \subfigure[Wireframe model, $\vec{x}_{\obs} = (14.422,0,2)^T$]{\includegraphics[width=0.497\textwidth]{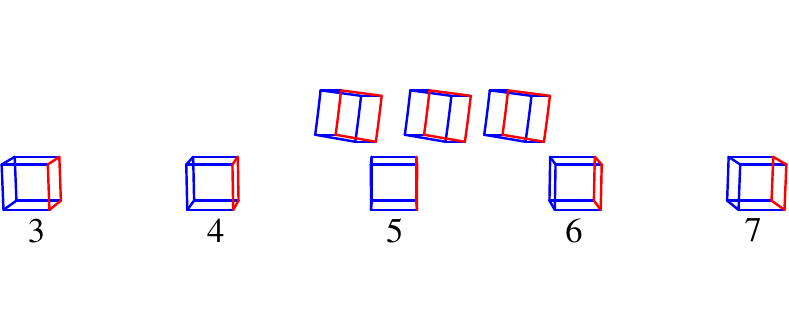}}
 \subfigure[Rendered, $\vec{x}_{\obs}=(8,12,2)^T$]{\includegraphics[width=0.497\textwidth]{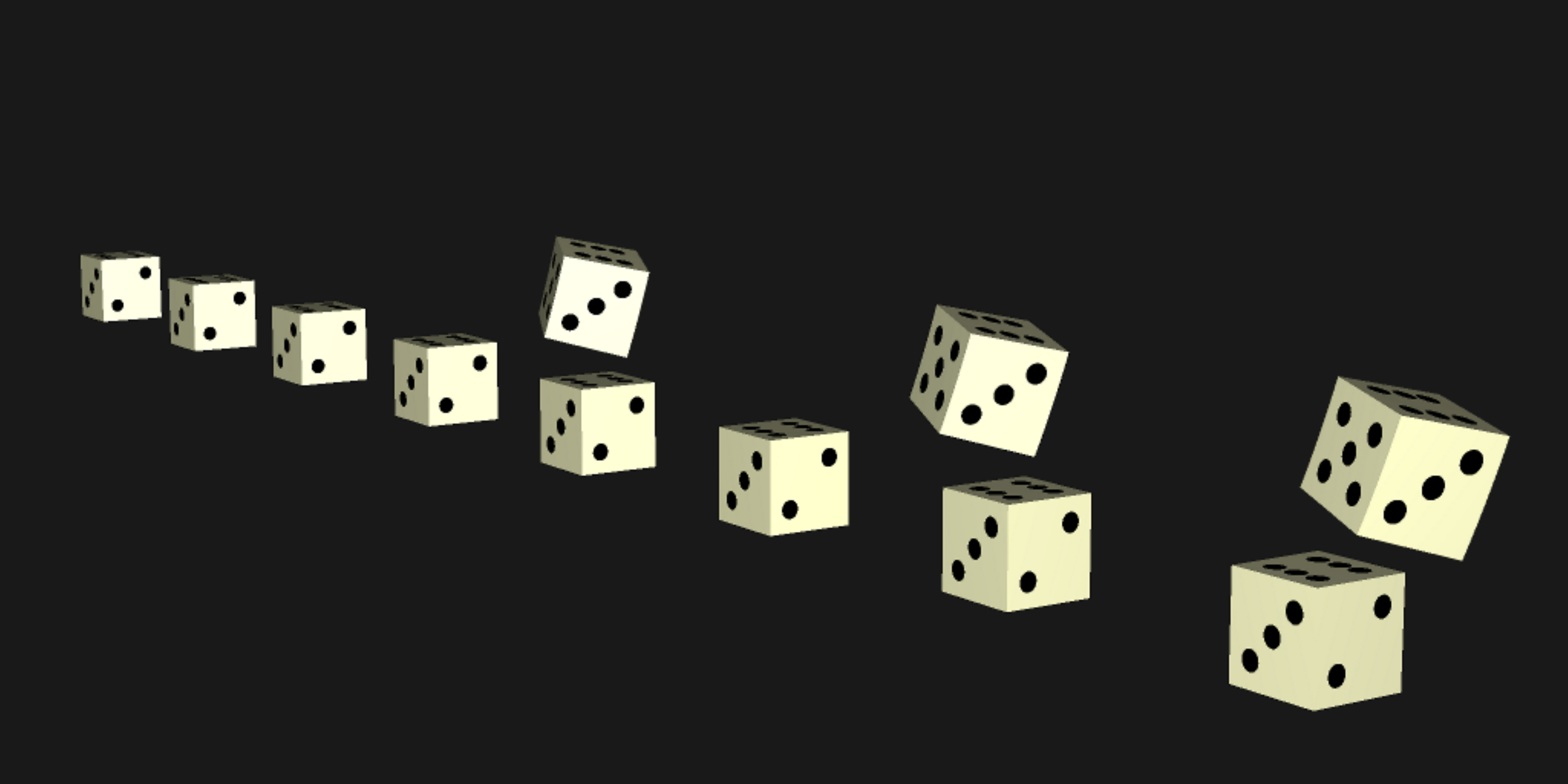}}\hfill
 \subfigure[Wireframe model, $\vec{x}_{\obs}=(8,12,2)^T$]{\label{fig:diceWF}\includegraphics[width=0.497\textwidth]{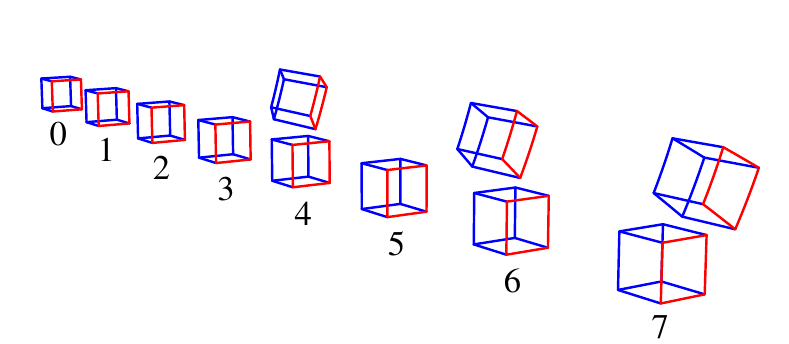}}
 \caption{Apparent view of a die with edge length $l'=0.5$ and velocity $\vec{\beta}=(0,0.9,0)^T$ moving above a row of static dice.
 The observer's pinhole camera has $32\degree\times 16\degree$ field of view. 
 The front edges of all die are coloured red to make their orientations easier to recognize.
 The observation times are $x_{\obs}^0=\left\{14.012,15.1,16.12\right\}$.
 In the top row, the observer is at  $\vec{x}_{\obs} = (14.422,0,2)^T$. 
 In the bottom row the observer has the same distance to the $x^3$-axis but she is located at $\vec{x}_{\obs}=(8,12,2)^T$,
 i.e. by an angle $\varphi = 56.3\degree$ shifted away from the $x^1$-axis.
 (Script: \texttt{appDie, animDie.py}). 
 } 
 \label{fig:dice}
\end{figure}

Consider a row of $8$ dice with edge length $l=0.5$ at rest in the reference frame $S$.
The centre of the $n$-th die is located at $\vec{x}_n = (0,-10+n\cdot 2,-0.75)^T$, $n=0,\ldots,7$.
Another die of the same size is at rest in the centre of the reference frame $S'$, while the frame $S'$ itself moves with velocity $\vec{\beta}=(0,0.9,0)^T$. 
An observer located at $\vec{x}_{\obs}$ looking into the direction of the origin of $S$ will see the row of dice and the moving die as shown in figure~\ref{fig:dice}.

Although the observation times for the moving die are equal in figure~\ref{fig:dice}, the apparent positions differ dependent on the position of the observer.
If she looks perpendicular to the row of dice, the light travel times are nearly the same.
But that is no longer true if the observer has a tilted view to the row.
Light from the rearmost position needs much more time than from a closer position. 
Hence, the distances between the apparent positions are longer.

% ----------------------------------------------
%   Apparent view of a circle/ball
% ----------------------------------------------
\subsection{Apparent view of a circle/ball}\label{subsec:expAppViewBall}
\begin{figure}[ht]
  \includegraphics[width=\textwidth]{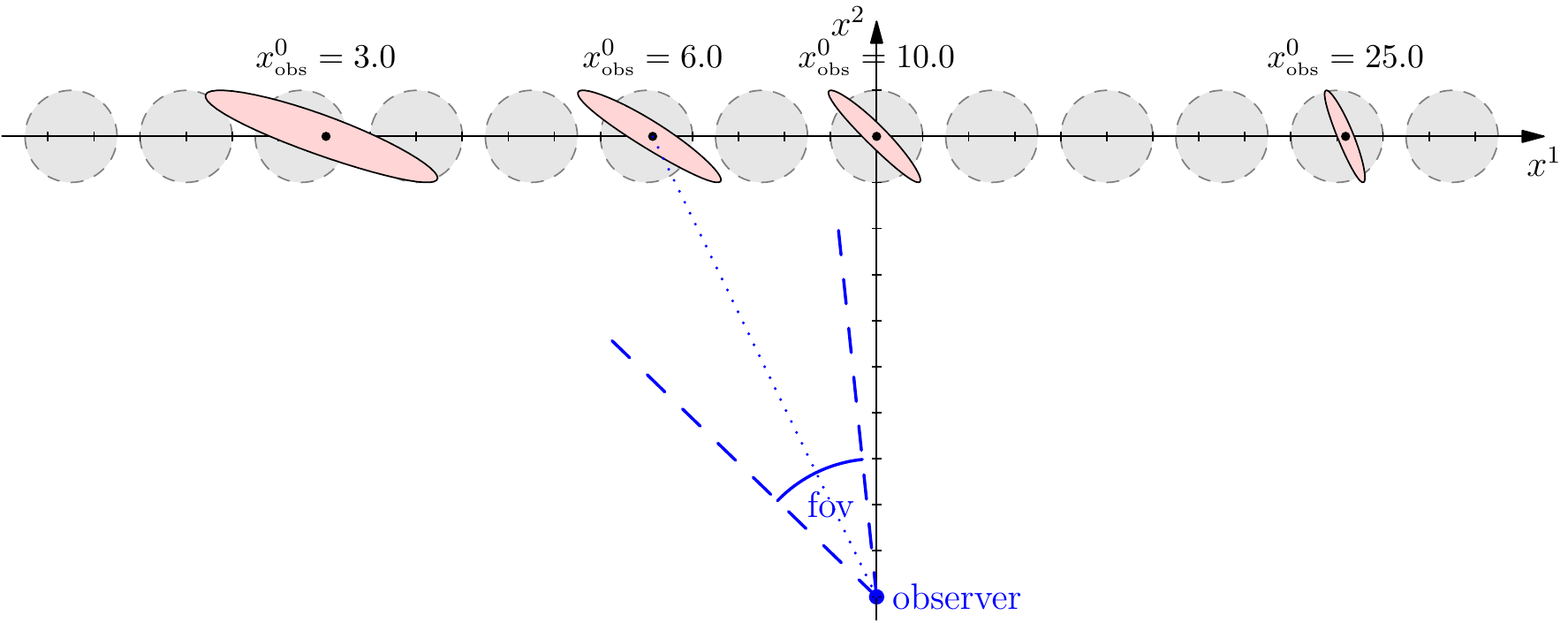}
  \caption{Apparent shapes (red) of a moving circle for an observer located at $\vec{x}_{\obs}=(0,-10,-1)^T$, an offset $\vec{a}=\vec{0}$, a velocity $\vec{\beta}=(0.9,0,0)^T$, and observation times $x_{\obs}^0$. 
  The central point of the circle (black dot) is at $\vec{x}'=(0,0,0)^T$ with respect to $S'$.
  The grey-dashed disks represent a row of static circles.
  (Script: \texttt{appCircle, animCircle.py})
  }
  \label{fig:appCircle}
\end{figure}

Penrose~\cite{Penrose:1959:Sphere} has already shown in 1959 that the apparent shape of a relativistically moving sphere is again a sphere.
However, the shape of the photo-object of the sphere is more similar to an ellipsoid, see figure~\ref{fig:appCircle}.
Hence, the observer will ``see'' an ellipse but the projection on his plane of sight leads to a circular outline and his brain therefore interprets it as a rotated sphere.

In order to follow the apparent image of a moving circle/ball, the camera has to point in the direction of the apparent position of the circle's/ball's centre $\vec{x}'=\vec{0}$.
In the standard configuration, $\vec{a}=\vec{0}$, $\vec{x}_{\obs}=(0,-y_{\obs},0)^T$, $\vec{\beta}=(\beta,0,0)^T$, we have $\rho=-x_{\obs}^0$ and $\vec{\eta}=-\vec{x}_{\obs}$.
Thus, the camera has to follow the apparent point $\vec{x}_{\cen}$,
\begin{equation}
  \vec{x}_{\cen} = \gamma\beta x'^0, \qquad\mbox{with}\quad x'^0 = \gamma x_{\obs}^0 - \sqrt{\gamma^2\beta^2(x_{\obs}^0)^2+y_{\obs}^2}.
\end{equation}

\begin{figure}[ht]
  \subfigure[Rendered]{\includegraphics[height=140px]{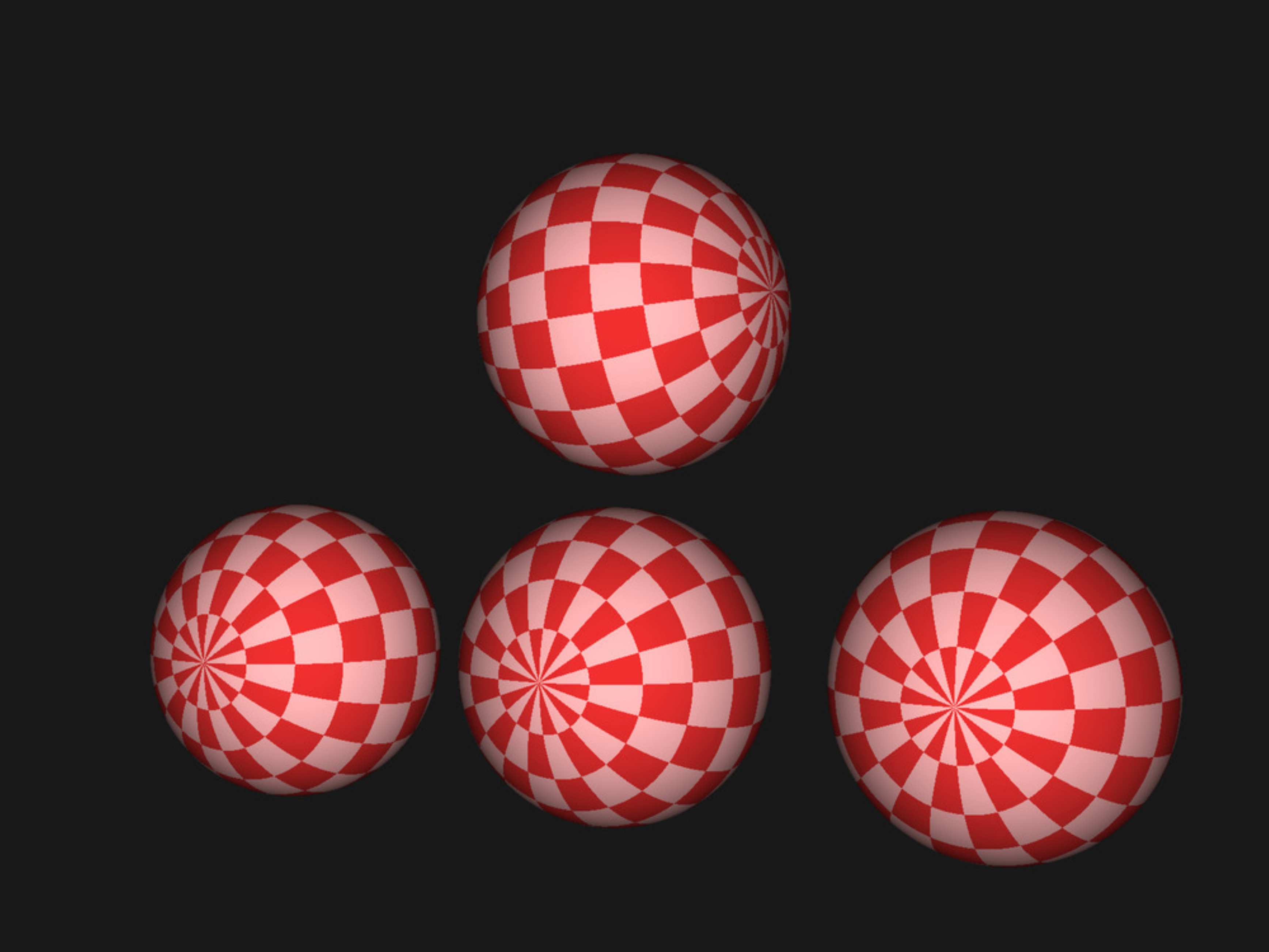}\label{fig:movSphereGeoViS}}\hfill 
  \subfigure[Wireframe model]{\includegraphics[height=140px]{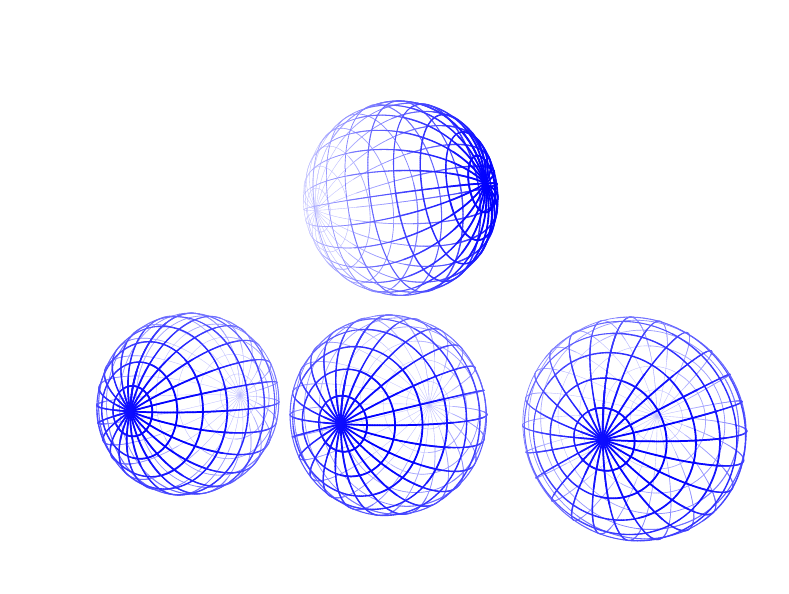} \label{fig:movSphereasy}}
  \caption{A chequered ball of radius $r'=1$ moves with $\beta=0.95$ along the $x^1$-axis in positive direction. 
  The lower balls are static and are positioned at $\vec{x}_n=(-17.5+n\cdot 2.5, 0, -2.2)^T$, $n=\{4,5,6\}$. 
  The observer is located at $\vec{x}_{\obs}=(0,-10,-1)^T$.
  The camera's field of view is $40\degree\times 30\degree$. 
  The inclusion of depth information in the wireframe picture reveals that the region around the moving sphere's right pole is closest to the observer as can also be seen from figure \ref{fig:appCircle} but which is hardly visible in the rendered image. 
  (Scripts: \texttt{appSphere} and \texttt{appSphereZ.asy} with depth information)}
  \label{fig:movSphere}
\end{figure}

Figure~\ref{fig:movSphere} shows a ball moving along the $x^1$-axis in positive direction above a row of static balls where the axes of all of them point in the same direction.
The image rendered using \emph{GeoViS} demonstrates clearly that the moving ball still appears as a ball but appears to be rotated only. 
If we visualize this situation with our standard wireframe model, we lose the spatial impression because the lines on the front and on the back of the sphere intersect on the plane of sight, resulting in a ``cluttered'' impression.
This of course is an intrinsic property of our wireframe models.
However, the transformation into the plane of view using the \emph{view} and \emph{projection} matrices preserves depth information in the $\hat{p}^3$-component of the projected point, see \ref{appsec:matrices} for a short discussion. 
We use this information to draw lines closer to the observer thicker and with stronger colours than lines further away. 
In fact, this can even help us to extract information that is not perceivable in the rendered images, namely that the right pole of the moving sphere is the part closest to the observer as we can already see in figure~\ref{fig:appCircle}.
This can nicely be seen in figure \ref{fig:movSphereasy} while it is not visible in the rendered image \ref{fig:movSphereGeoViS}.
However, the scripts that  produce these figures are significantly more complicated, as we have to subdivide the picture in small line segments, sort them with respect to their depth value
and draw them in depth-ascending order.
Therefore, we also include simpler scripts that do not use depth information but which might be easier to read.

% ----------------------------------------------
%   Apparent view for close encounters
% ----------------------------------------------
\subsection{Apparent view for close encounters}
In our previous examples the distortion effects due to the finite speed of light are relatively small, because the distance of the observer to the objects is large in comparison to their size.
If the observer's distance is comparable to the object's scale, the time of flight for light rays originating from different locations on the object's surface varies strongly. 
Hence, the observer sees different regions of the object at very different times and therefore locations and so the object appears strongly distorted.

% ----------------------------------------------
%  Apparent view of a line/rod
% ----------------------------------------------
\subsubsection{Apparent view of a line/rod oriented perpendicular to its direction of motion}\label{subsex:expAppViewLinePerp}

We again start with the discussion of a moving rod, but contrary to section \ref{subsec:expAppViewLineAlong} we now assume it to be aligned perpendicularly to its direction of motion.
In this case, the apparent view becomes more interesting.
Let $\vec{x}_{\obs}=\vec{a}=\vec{0}=\vec{x}'$, $\vec{\beta}=(\beta,0,0)^T$, and $\vec{\sigma}'=(0,1,0)^T$.
Then, $\vec{\beta}\cdot\vec{\sigma}'=0$,  $\rho=-x_{\obs}^0$, $\vec{\eta}=\vec{0}$, and $\upsilon=0$.
Furthermore, $\omega_{\lin}(s')^2=\gamma^2\beta^2(x_{\obs}^0)^2+s'^2$.

\begin{figure}[ht]
\centering
  \includegraphics[scale=0.8]{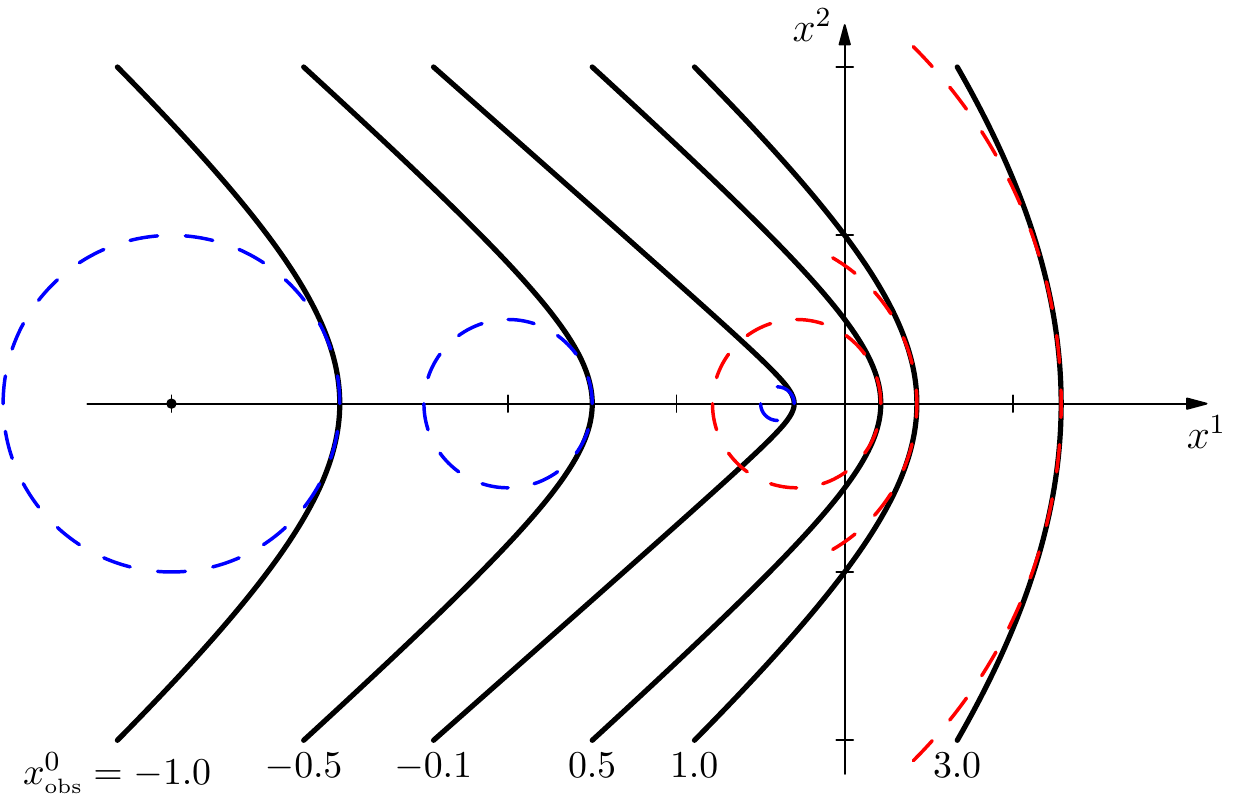}
  \caption{A rod of length $l'=4$ is oriented perpendicularly to its direction of motion, $\vec{\beta}\cdot\vec{\sigma}'=0$.
   Here, $\beta=0.75$. 
   The observer is located at $\vec{x}_{\obs}=\vec{0}$ and $\vec{a}=\vec{0}$.
   The osculating circle has radius $(\kappa^{\perp})^{-1}=|x_{\obs}^0|$.
   (Script: \texttt{appRod, animRod.py})
   }
   \label{fig:hyperbola}
\end{figure}

\begin{figure}[ht]
  \centering
  \includegraphics[scale=0.8]{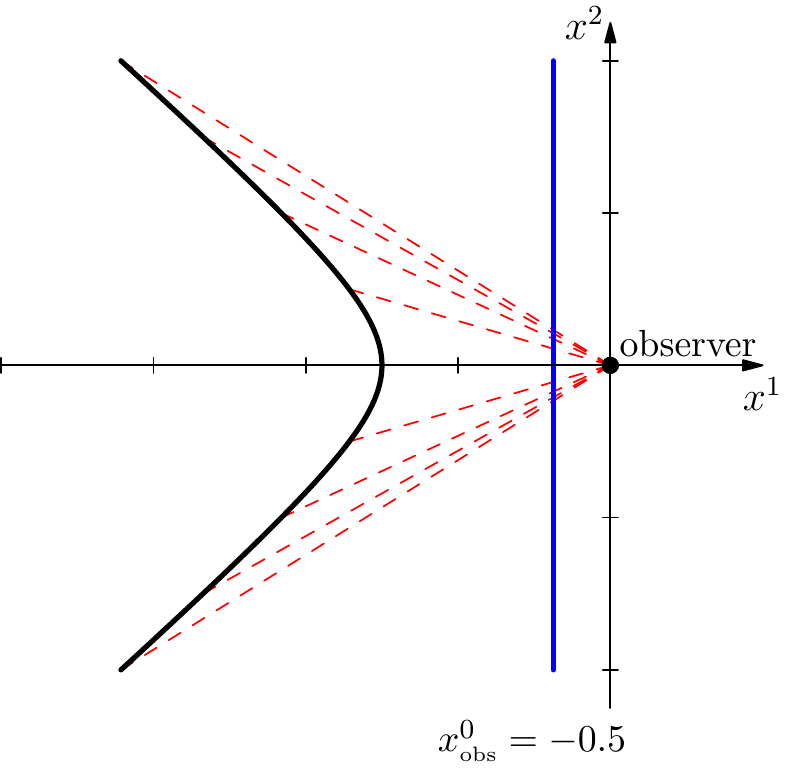}
  \caption{Photo-object for a rod with length $l'=4$ moving with $\beta=0.75$ towards the observer. 
  At observation time  $x^{0}_{\obs}=-0.5$, the observer receives light rays from the outer parts of the rod that started already at approximately $x^{0}_{\obs}=-3.0$.
   (Script: \texttt{appRodLight.asy, animRodLight.py})
   }
   \label{fig:appRodLight}
\end{figure}

\begin{figure}[ht]
  \hspace{0.25\textwidth}\subfigure[Polygon rendering]{\includegraphics[height=80px]{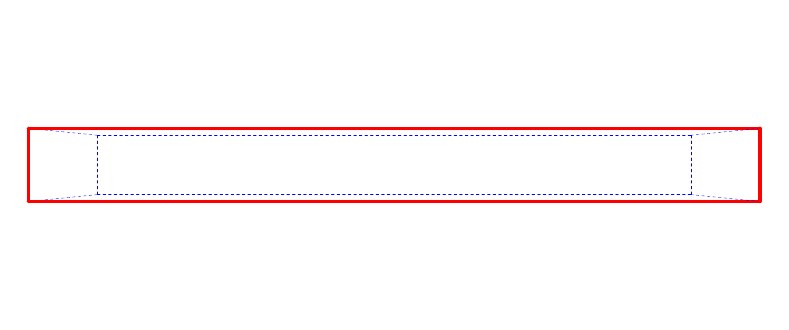} \label{fig:movingRodPolygonrendering}}
  
  \subfigure[Rendered]{\includegraphics[height=80px]{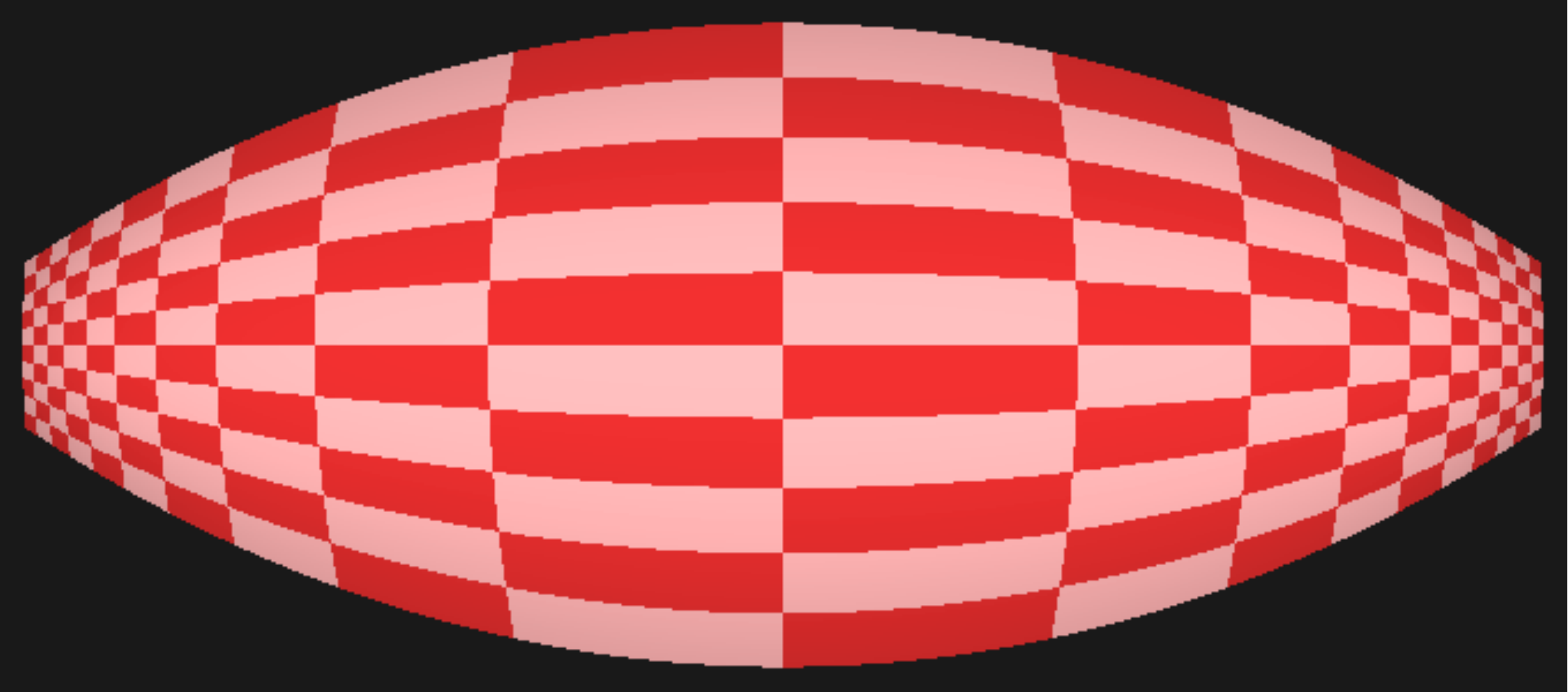}\label{fig:movingRodGeoVis} } \hfill
  \subfigure[Wireframe model]{\includegraphics[height=80px]{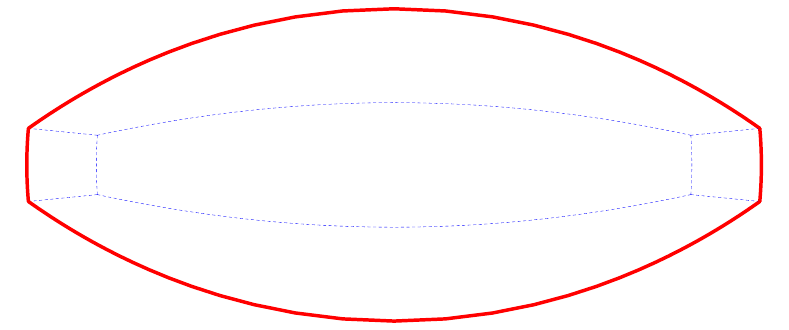} \label{fig:movingRodasy}}
  \caption{A rod moves with $\beta=0.9$ along the positive $x^1$-direction. 
  The observation time is given by $x_{\obs}^0=-0.104$ and the observer looks along the negative $x^1$-direction.   
  The front edges of the rod are coloured red to make its orientation easier to recognize.
  (Script: \texttt{appRodView, animRodView.py})
  }
  \label{fig:movingRod}
\end{figure}

As already discussed in section~\ref{subsec:appPosLine}, the perpendicularly oriented line appears as a hyperbola which can be described by the implicit equation
\begin{equation}
  1 = \frac{(x^1-\gamma^2\beta x_{\obs}^0)^2}{\gamma^4\beta^4(x_{\obs}^0)^2} - \frac{(x^2)^2}{\gamma^2\beta^2(x_{\obs}^0)^2}.
\end{equation}
The apex resides on the $x^1$-axis with curvature $\kappa^{\perp}(s'=0)=1/x_{\obs}^0$, see equation~(\ref{eq:curvPerp}).
The centre of the osculating circle has coordinates $x^1=\gamma^2(\beta x_{\obs}^0-|x_{\obs}^0|)$ and $x^2=0$, see figure~\ref{fig:hyperbola}.
At $x_{\obs}^0=0$, the hyperbola degenerates to a corner.
As light rays originating from points close to the middle of the rod take much less time to reach the observer than those from its outer parts, the observer sees the outer parts at earlier times and, hence, at larger distances than the centre and the rod appears to be bent. 
Figure~\ref{fig:appRodLight} illustrates this situation for a rod moving with $\beta=0.75$. 
At observation time $x^{0}_{\obs}=-0.5$, were $x^{0}_{\obs}=0$ is defined as the time when the rod reaches the observer, 
the moving rod is already very close to the observer, but the light rays from its outer parts left its surface as early as approximately $x^{0}_{\obs}=-3.0$ and therefore the observer gets the impression that the rod is still quite far away.
Figure~\ref{fig:movingRod} shows a moving rod which is described by a cuboid with lower left corner $\vec{c}'_{\lleft}=(-0.1,-1.0,-0.1)^T$ and upper right corner $\vec{c}'_{\uright}=(0.1,1.0,0.1)^T$.
The longitudinal direction is oriented along the $x^2$-axis and the rod moves along the positive $x^1$-direction. 
In figure \ref{fig:movingRodPolygonrendering} we show an example of polygon rendering, where only
the rod's vertices are transformed to their apparent positions. 
This technique obviously is insufficient to correctly visualize situations where strong distortions appear, because the edges connecting the vertices remain straight lines, for a more detailed discussion see \cite{Mueller:2010:LRT}.
Figures \ref{fig:movingRodGeoVis} and \ref{fig:movingRodasy} compare the results of the four-dimensional ray tracing with \emph{GeoViS} and of our wireframe model \emph{asymptote} script. 
As we transform the entire edges and not just the vertices, their hyperbolic shape becomes apparent. 

% ----------------------------------------------
%   Sphere and Cube in close fly by
% ----------------------------------------------
\subsubsection{Sphere and Cube in close fly by}

\begin{figure}[ht]
  \subfigure[Cube at rest]{\includegraphics[width=0.495\textwidth]{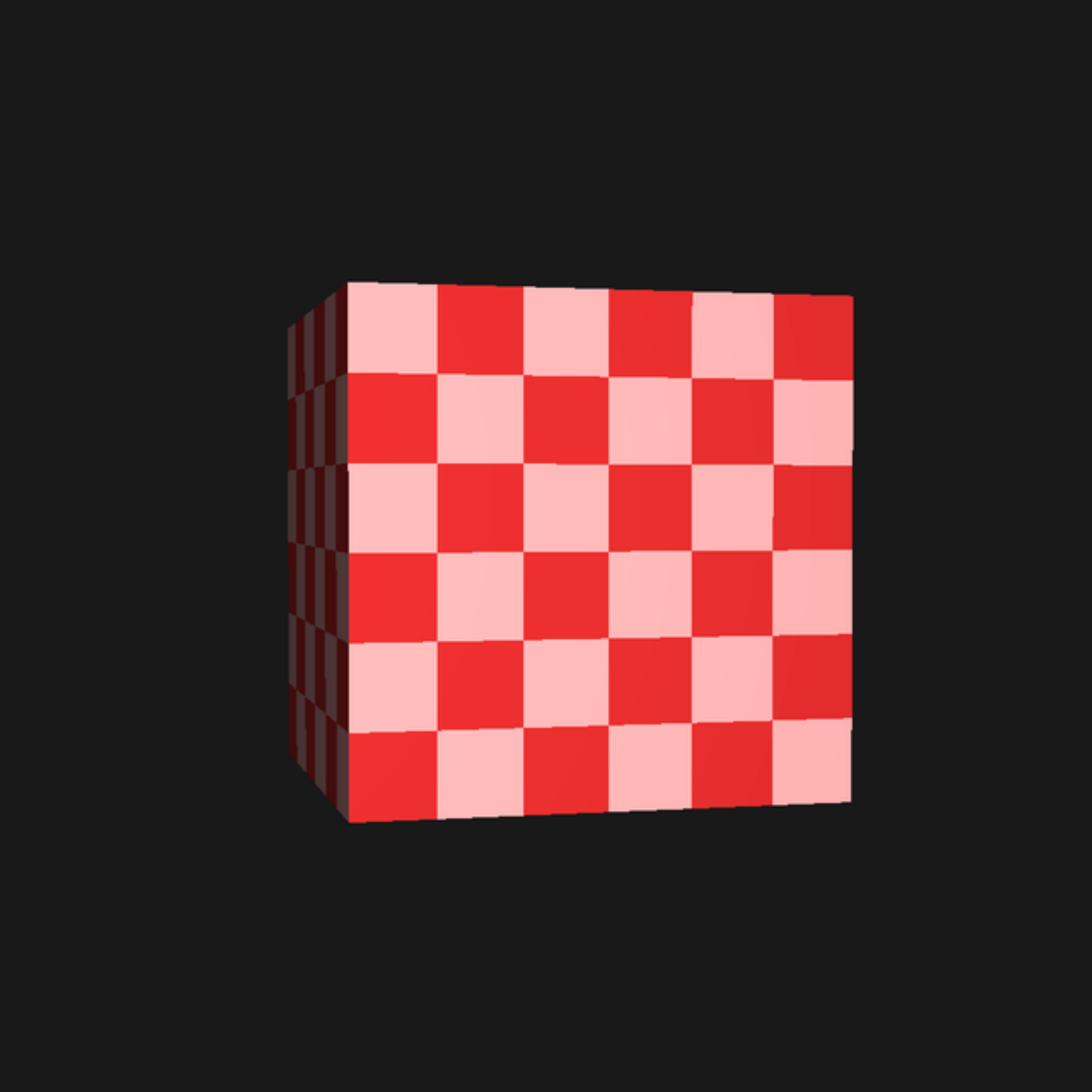}}\hfill
  \subfigure[Cube with $\beta=0.9$]{\includegraphics[width=0.495\textwidth]{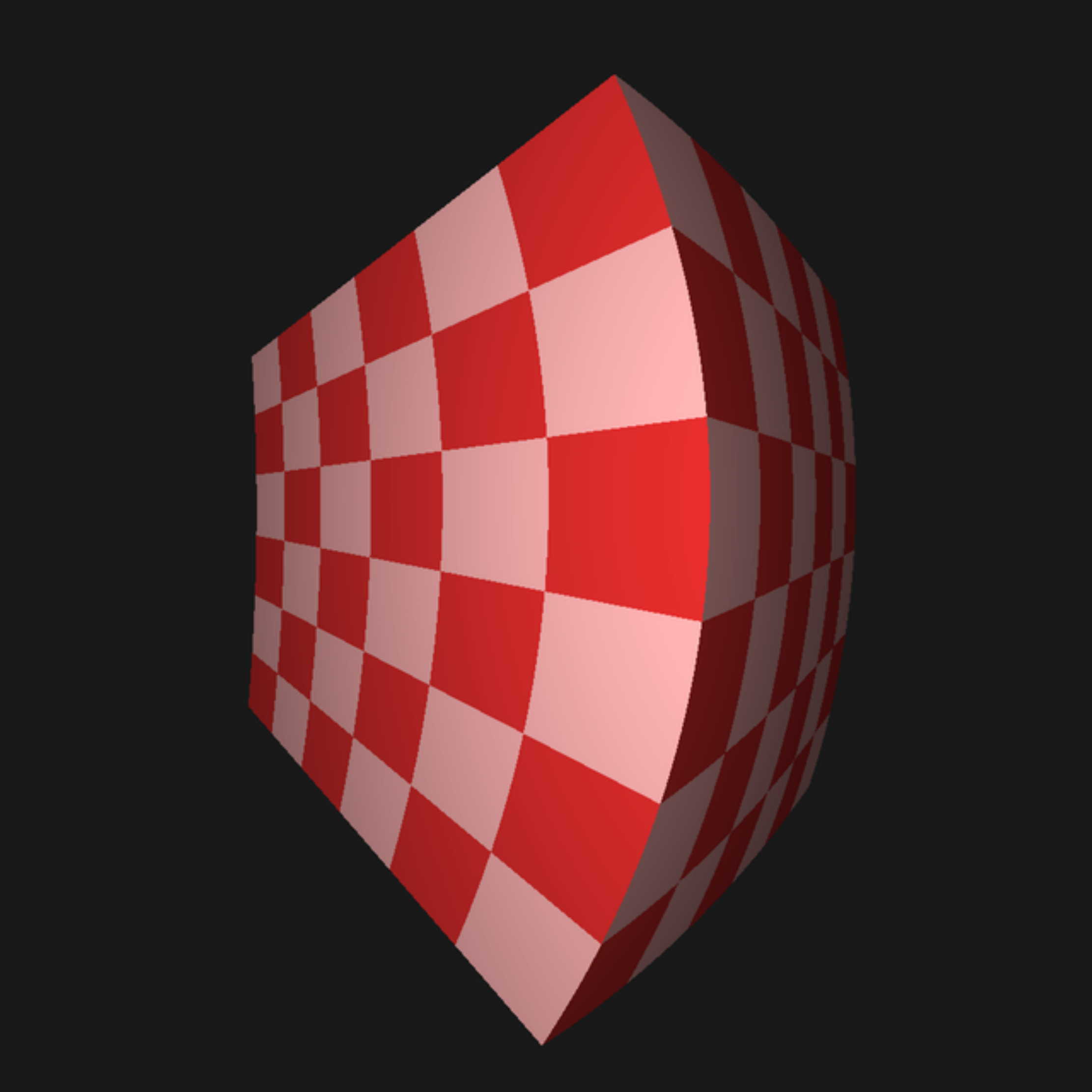}}\\
  \subfigure[Cube at rest]{\includegraphics[width=0.495\textwidth]{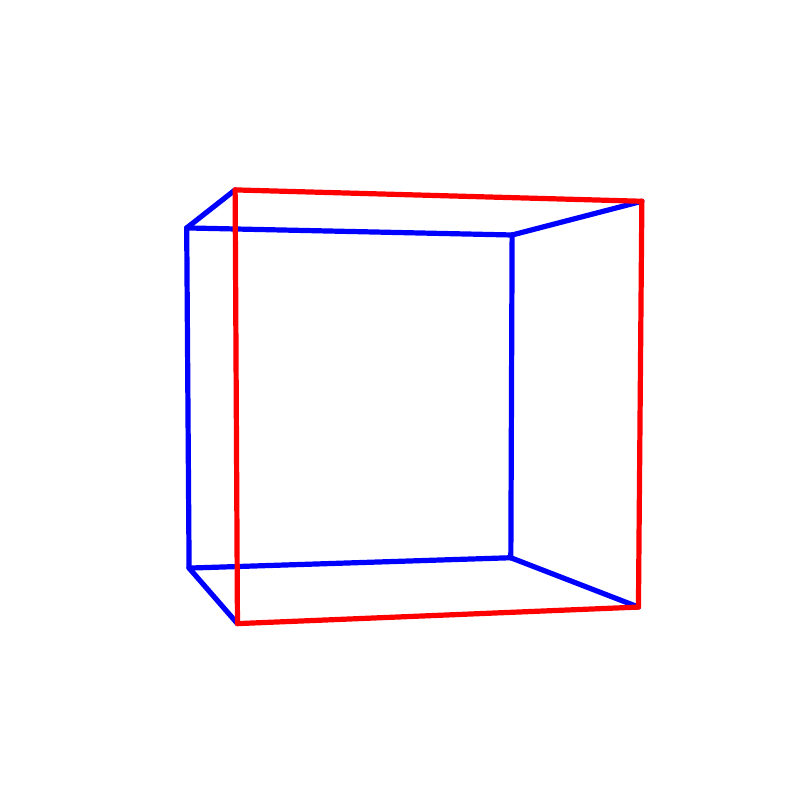}}\hfill
  \subfigure[Cube with $\beta=0.9$]{\includegraphics[width=0.495\textwidth]{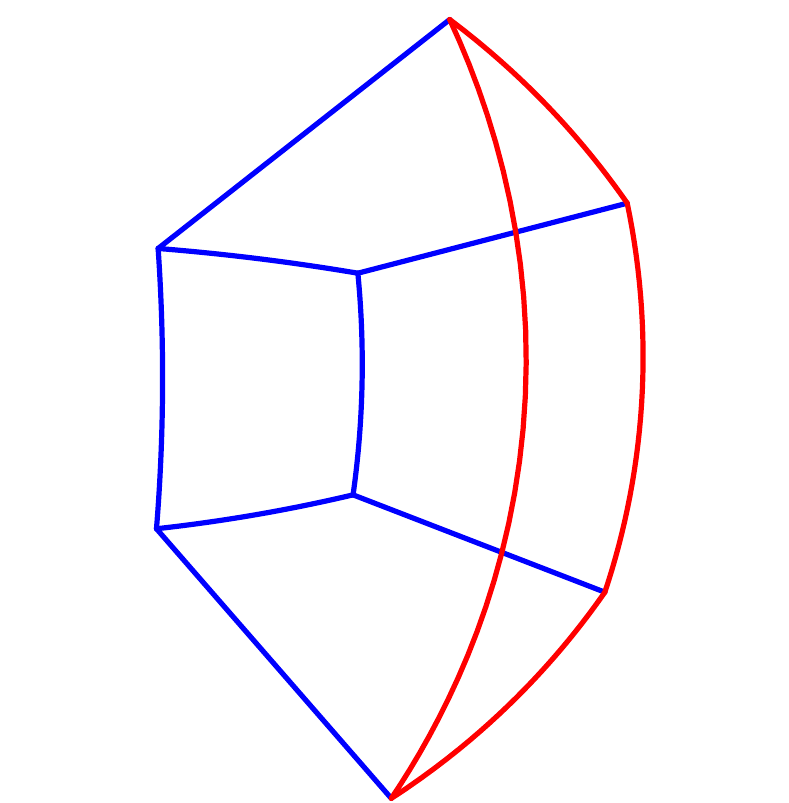}}  
  \caption{Apparent distortion of a cube with edge length $l'=0.5$ in close fly by. 
  The camera has a field of view of $32\degree\times 32\degree$.
  The observer is at position $\vec{x}_{\obs}=(0.5,2,0.05)^T$, i.e. at a distance $d=1.5$ to the $x^3$-axis, and the cube moves along the positive $x^2$-direction. (Script: \texttt{appCube})
  }
  \label{fig:closeflybyCube}
\end{figure}

\begin{figure}[ht]
  \subfigure[Sphere at rest]{\includegraphics[width=0.495\textwidth]{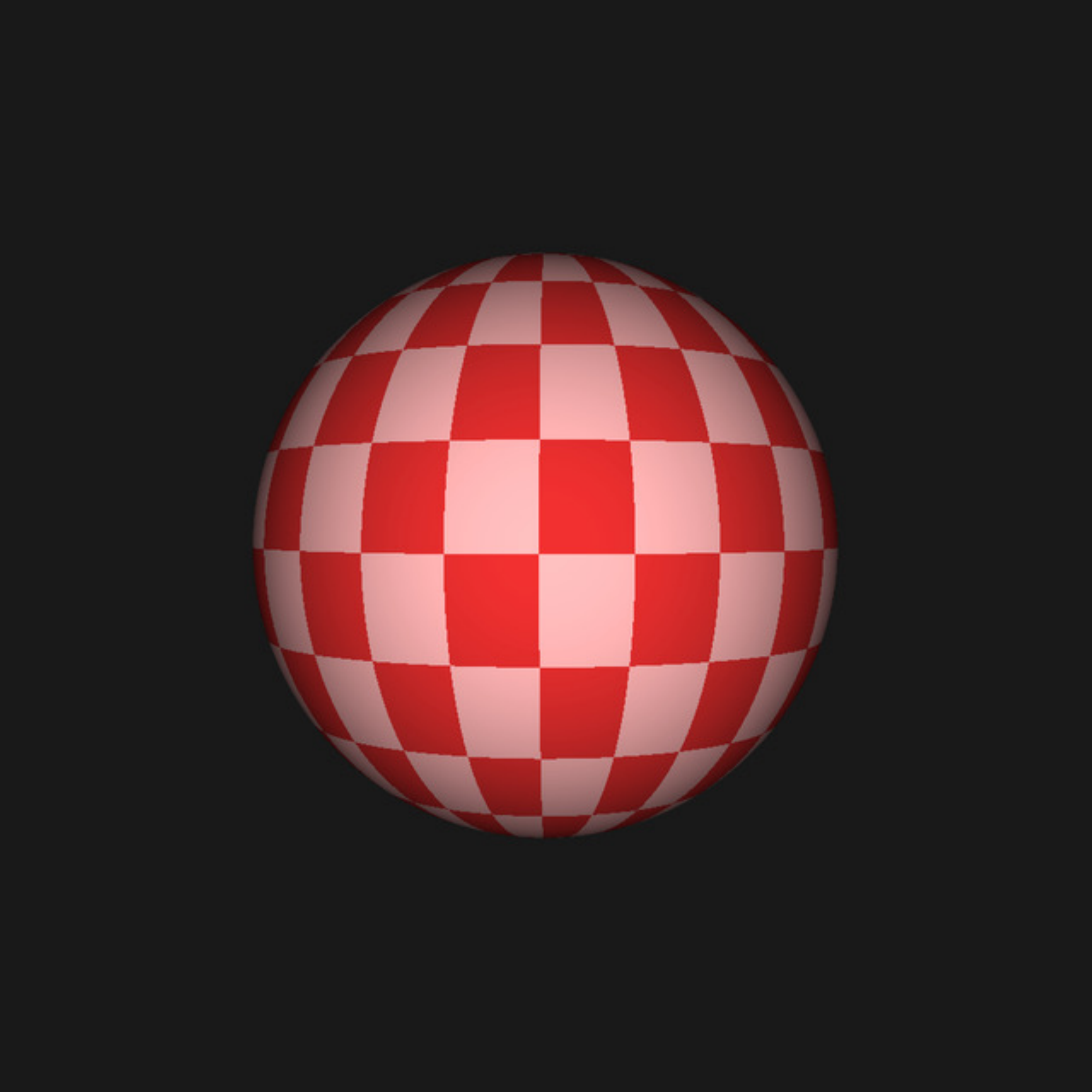}}\hfill
  \subfigure[Sphere with $\beta=0.9$]{\includegraphics[width=0.495\textwidth]{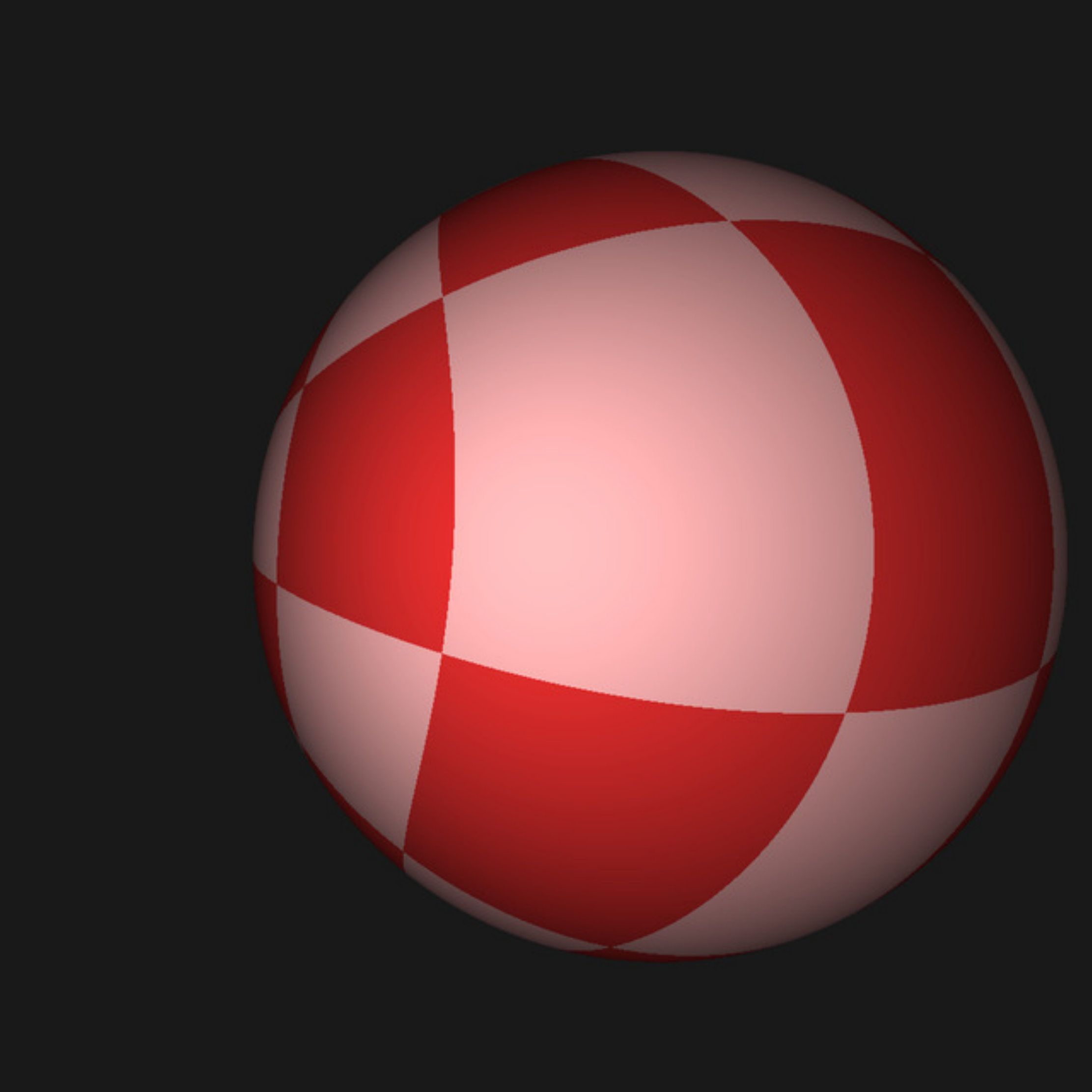}}\\
  \subfigure[Sphere at rest]{\includegraphics[width=0.495\textwidth]{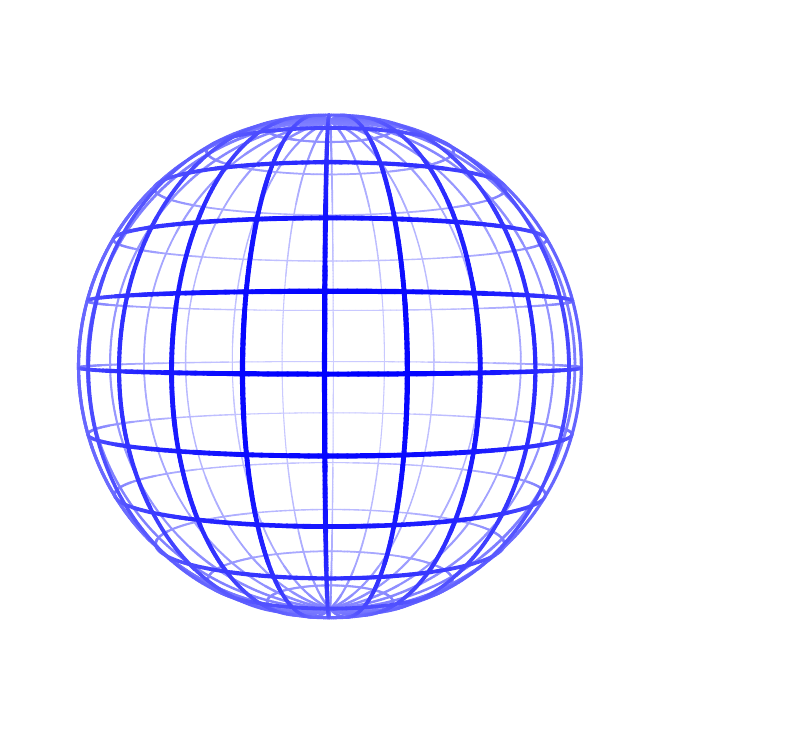}}\hfill
  \subfigure[Sphere with $\beta=0.9$]{\includegraphics[width=0.495\textwidth]{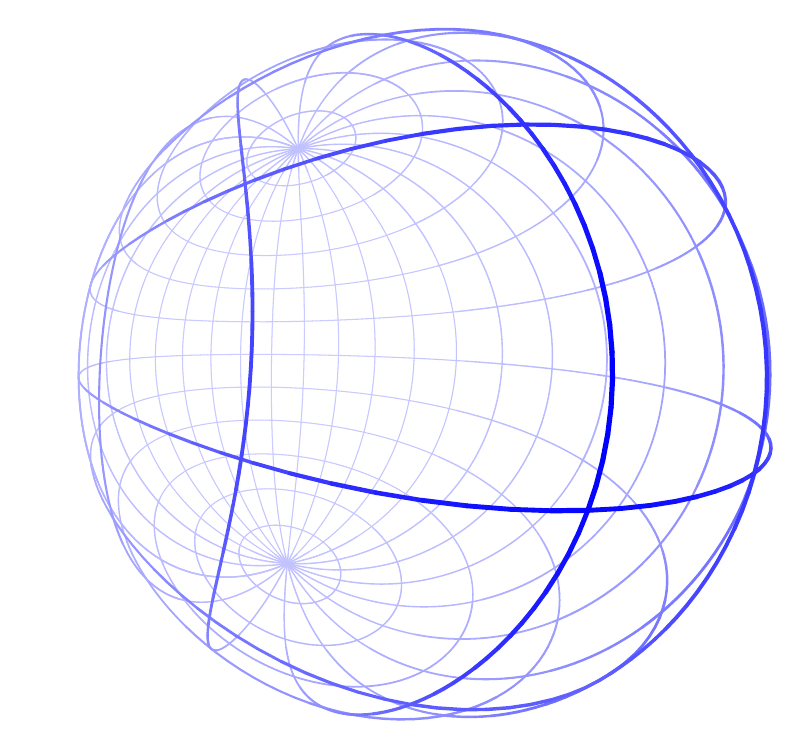}}
  \caption{Apparent distortion of a sphere with radius $r'=0.5$ in close fly by. 
  The camera has a field of view of $50\degree\times 50\degree$.
  The observer is at position $\vec{x}_{\obs}=(0.5,2,0.05)^T$, i.e. at a distance $d=1.5$ to the $x^3$-axis, and the sphere moves along the positive $x^2$-direction. (Scripts: \texttt{appSphereSingle} and \texttt{appSphereSingleZ.asy} with depth information)  
  }
  \label{fig:closeflybySphere}
\end{figure}

We conclude our examples with a comparison of a cube and a sphere at rest with their moving counterparts closely passing the observer, see figures~\ref{fig:closeflybyCube} and \ref{fig:closeflybySphere}.

In both cases the scene is chosen such that the apparent centre of the moving object coincides with the centre of the static object.
Contrary to the rod example, these cases are not symmetric because the objects are not moving towards the observer. 

In the cube example, the different appearances of lines oriented perpendicularly or parallely to their direction of motion becomes quite apparent. 
The upper and lower edges of the cube are oriented almost parallely to the direction of motion and hence appear straight. 
On the other hand the edges of the back and the front are oriented almost perpendicularly to the direction of motion and appear bent.
This effect is much stronger for the front of the cube than for its back, because these edges are close to the observer and the flight times for the light rays from different points on these edges differ more strongly.

\begin{figure}[ht]
  \centering
  \includegraphics[width=0.9\textwidth]{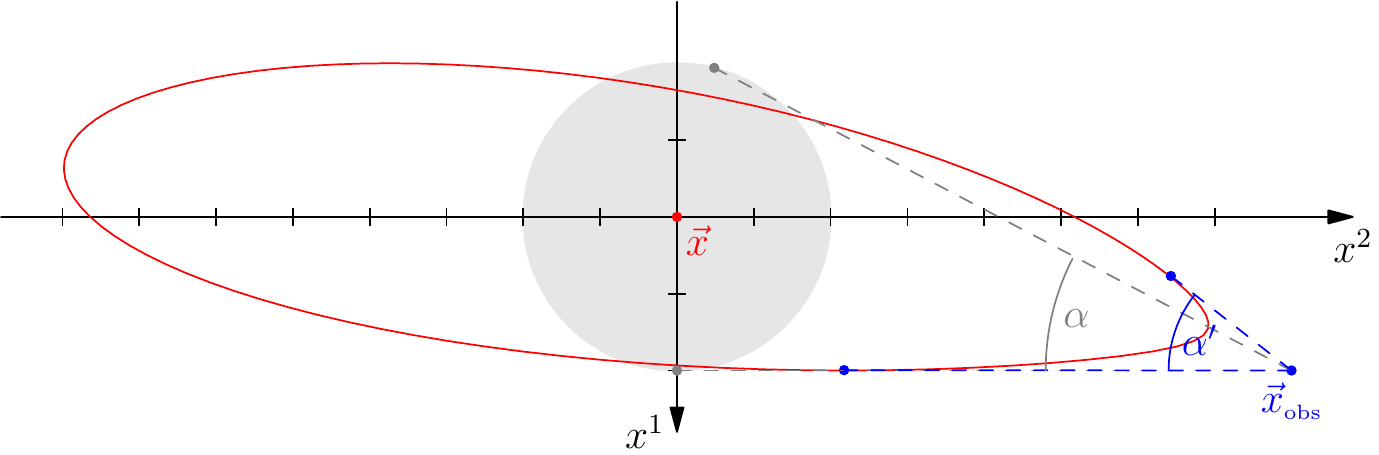}
  \caption{Two-dimensional analog of figure~\ref{fig:closeflybySphere}. 
  The angular size $\alpha'$ of the apparent sphere, represented by the red photo-object, is greater than the angular size $\alpha$ of the static sphere (gray disk) at the apparent position of the sphere's centre, here $\vec{x}=\vec{0}$.}
  \label{fig:appSizeOfSphere}
\end{figure}

The sphere example clearly shows that the sphere retains its circular shape while its surface is strongly distorted, in accordance with the results by Penrose~\cite{Penrose:1959:Sphere}.
However, it appears larger than the sphere at rest, see also figure~\ref{fig:appSizeOfSphere}. Please note again that the centres of the moving sphere and the one at rest coincide so this is indeed a, well-known, relativistic effect.
This example also again impressively demonstrates, how we can enhance the visual impression by including depth information.

% -----------------------------------------------------------------
%    Summary
% -----------------------------------------------------------------
\section{Summary}
In this article we derived general equations that describe the apparent view of relativistically moving points, lines, and spheres.
We implemented our results in \emph{asymptote} and \emph{python} scripts and generated some exemplary scenes of wireframe objects and compared our results for these cases with images created with a four-dimensional ray tracer.
We showed that by taking into account the depth information, our wireframe figures can provide a realistic impression of the special relativistic distortion effects. 
The tools that we created are very flexible and may be used to study other scenes, while the examples that we created can already serve as an aid in teaching of the visual appearance of relativistically moving objects.

% -----------------------------------------------------------------
%    appendix
% -----------------------------------------------------------------
\appendix

% -----------------------------------------------------------------
%    Further examples
% -----------------------------------------------------------------
\section{Further examples}\label{appsec:examples}
In the following, we will give some additional examples that could be used directly in the classroom either for demonstration purposes or as exercises.

\begin{srExercise}
Given a rod of length $l'=4$ which moves perpendicularly to its orientation towards an observer, see Figs.~\ref{fig:hyperbola} and \ref{fig:appRodLight}. 
Play around with the velocity $\beta$ and explain why the rod appears to be bent stronger the faster it moves.
\end{srExercise}

\begin{srResult}
When the velocity of the rod comes ever closer to the speed of light, the light travel times from the different positions of the rod to the observer become more and more diverse. 
Thus, light from the top of the rod has to start ever earlier than light from the center of the rod in order to reach the observer at the same time which results in an increasing bending of the rod.
\end{srResult}

\begin{srConf}
  demoRodLight.py
\end{srConf}
\vspace*{0.3cm}

\begin{srExercise}
Given a sphere of radius $r'=0.5$ moving along the positive $x^1$-direction with velocity $\beta$. 
The observer is located at $\vec{x}_{\obs}=(0,-100,0)^T$ and looks along the $x^2$-axis, compare Fig.~\ref{fig:appCircle}.
At fixed observation time $x_{\obs}^0=100$, he will see that the sphere is apparently rotated in its direction of motion.
Determine the relation between the sphere's velocity $\beta$ and the apparent rotation angle $\alpha$.
\end{srExercise}

\begin{srResult}
For $\beta=0$, the axis/pole of the sphere points towards the observer.
With increasing velocity, the sphere appears to be rotated by an angle $\alpha=\arctan(\beta\gamma)$. 
(Terrell~\cite{Terrell:1959:InvLC} uses the complementary angle.)
The angle $\alpha$ can be read from the image generated by the script. 
Given the distance $d$ of the pole to the center of the sphere and the radius $r$, both in relative units or pixels, the angle reads $\alpha=\arcsin(d/r)$.
\end{srResult}

\begin{srConf}
  demoSphere.py
\end{srConf}

\pagebreak
\begin{srExercise}
Analogous to Fig.~\ref{fig:diceWF}, a die moves with $\beta=0.9$ above a row of static dice.
Here, we fix the observation times to $x_{\obs}^0=\{13.422,14.422,15.422\}$ and let the observer rotate around the point of interest $(0,0,2)^T$ on the circle $\vec{x}_{\obs}=(r_{\obs}\cos\varphi,r_{\obs}\sin\varphi,2.0)^T$ with $r_{\obs}=14.422$ and $0\leq\varphi\leq 2\pi$.
Explain why the distances between the apparent positions of the moving die for the different observation times depend on the angle of observation $\varphi$.
What happens if $\beta$ is changed?
\end{srExercise}

\begin{figure}[h!]
  \centering
  \includegraphics[width=0.6\textwidth]{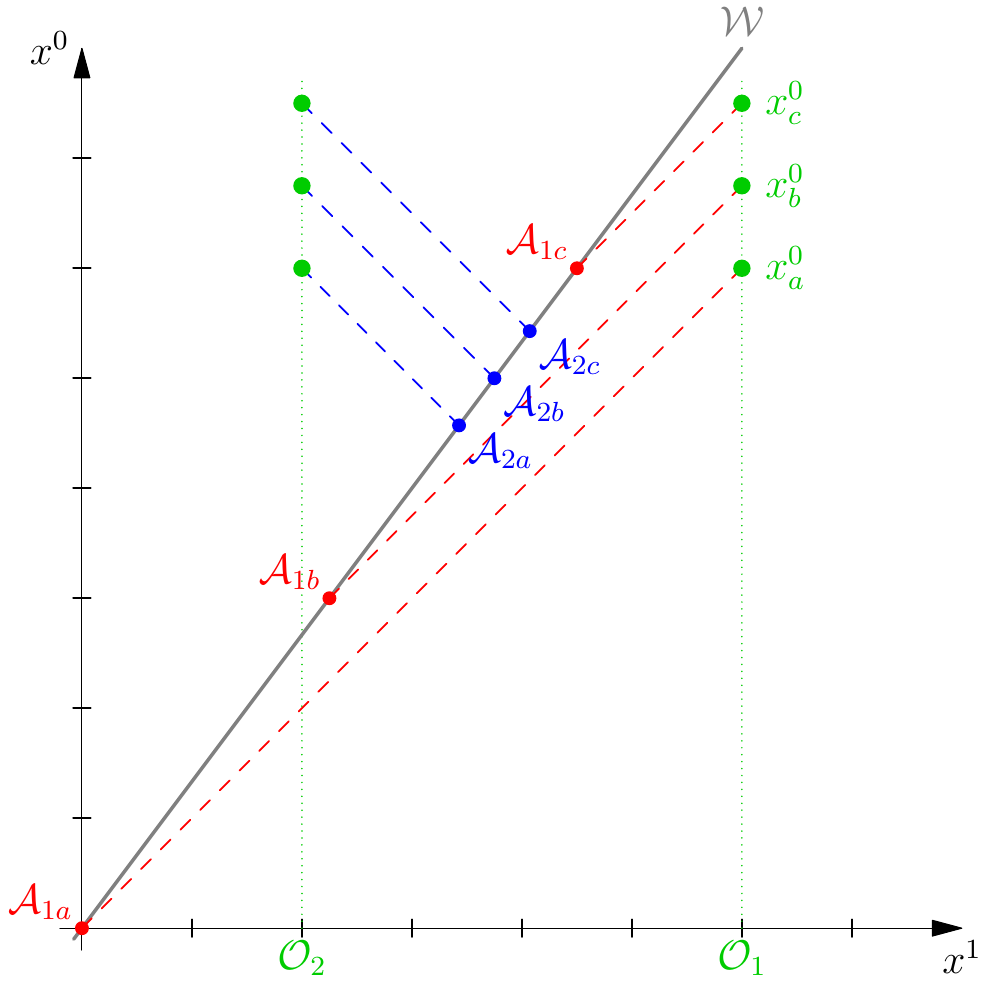}
  \caption{A point-like object, e.g. the center of the die, moves with $\beta=0.75$ in the positive $x^1$-direction, here indicated by the gray worldline $\mathcal{W}$ in the Minkowski diagram.
  The observation times for the static observers $\mathcal{O}_1$ and $\mathcal{O}_2$ are the same, $x_a^0<x_b^0<x_c^0$.
  The dashed lines represent parts of the backward light cones for the corresponding observers and observation times.
  The distances between the apparent positions $\mathcal{A}_{1i}$ for the approaching point are bigger than the distances of the apparent positions $\mathcal{A}_{2i}$ for the receding point; for example $|x^1(\mathcal{A}_{1a})-x^1(\mathcal{A}_{1b})| > |x^1(\mathcal{A}_{2a})-x^1(\mathcal{A}_{2b})|$
  }
  \label{fig:minkAppPos}
\end{figure}

\begin{srResult}
The observation time $x_{\obs}^0=14.422$ is chosen such that the apparent position of the die keeps its position irrespective of the observation angle $\varphi$ as long as $r_{\obs}=14.422$.
If $\varphi=\pi/2$ or $\varphi=3\pi/2$, the observer looks along the row of dice towards the approaching or receding die, respectively. 
Then, the finite speed of light has strong influence on where the moving die appears.
This can be most easily understood by means of a Minkowski diagram, see Fig.~\ref{fig:minkAppPos} for a similar situation with only a point-like object.
If $\varphi=0$ or $\varphi=\pi$, light travel times from the current positions of the moving die to the observer are nearly the same.
Hence, the distances between the apparent positions approximately reflect the actual distances between the current positions for the different observation times.
\end{srResult}

\begin{srConf}
  demoDie.py
\end{srConf}

% -----------------------------------------------------------------
%     Circular silhouette of a moving sphere
% -----------------------------------------------------------------
\section{Circular silhouette of a moving sphere}\label{appsec:silhouette}
As proven already by several authors, see for example Penrose~\cite{Penrose:1959:Sphere} or Boas~\cite{Boas:1961:LORS}, the silhouette of a relativistically moving sphere keeps circular irrespective of its velocity.
To show this circular silhouette for our general standard configuration, we could follow two approaches.

% -----------------------------------------------------------------
\subsection{Straightforward calculation}
The straightforward approach works as follows.
First, we determine the normal vector $\vec{n}$ at each apparent point $\vec{x}$ which is given by the cross product between the derivatives of $\vec{x}$ with respect to $\vartheta'$ and $\varphi'$, respectively,
\begin{equation}
   \vec{n}=\frac{\partial\vec{x}}{\partial\vartheta'}\times\frac{\partial\vec{x}}{\partial\varphi'}.
\end{equation}
For that, we need the derivatives
\begin{equation}
  \frac{\partial x'^0}{\partial\vartheta'} = -\frac{\partial\omega_{\sph}}{\partial\vartheta'},\quad 
  \frac{\partial x'^0}{\partial\varphi'} = -\frac{\partial\omega_{\sph}}{\partial\varphi'},\quad
  \frac{\partial\vec{\mu}}{\partial\vartheta'} = \frac{\partial\vec{\mu}}{\partial\varphi'} = \vec{0},
\end{equation}
where
\begin{equation}
  \frac{\partial\omega_{\sph}}{\partial\vartheta'} = \frac{1}{\omega_{\sph}}\vec{\mu}\cdot\sum\limits_{i=1}^3\frac{\partial s'_i}{\partial\vartheta'}\vec{\sigma}'_i\quad\mbox{and}\quad \frac{\partial\omega_{\sph}}{\partial\varphi'} = \frac{1}{\omega_{\sph}}\vec{\mu}\cdot\sum\limits_{i=1}^3\frac{\partial s'_i}{\partial\varphi'}\vec{\sigma}'_i.
\end{equation}
Therefrom, we obtain
\begin{eqnarray}
  \frac{\partial\vec{x}}{\partial\vartheta'} &=& \left[\sum\limits_{i=1}^3\frac{\partial s'_i}{\partial\vartheta'}\vec{\sigma}'_i\cdot\left(-\frac{\gamma}{\omega_{\sph}}\vec{\mu}+\frac{\gamma^2}{\gamma+1}\vec{\beta}\right)\right]\vec{\beta} 
  + \sum\limits_{i=1}^3\frac{\partial s'_i}{\partial\vartheta'}\vec{\sigma}'_i, \label{eq:dxdtheta} \\
  \frac{\partial\vec{x}}{\partial\varphi'} &=& \left[\sum\limits_{i=1}^3\frac{\partial s'_i}{\partial\varphi'}\vec{\sigma}'_i\cdot\left(-\frac{\gamma}{\omega_{\sph}}\vec{\mu}+\frac{\gamma^2}{\gamma+1}\vec{\beta}\right)\right]\vec{\beta} 
  + \sum\limits_{i=1}^3\frac{\partial s'_i}{\partial\varphi'}\vec{\sigma}'_i.\label{eq:dxdphi}
\end{eqnarray}
When building the cross product of (\ref{eq:dxdtheta}) and (\ref{eq:dxdphi}), we can make use of $\vec{\beta}\times\vec{\beta}=\vec{0}$ and the orthonormality of the basis vectors, $\vec{\sigma}'_i\times\vec{\sigma}'_j=\epsilon_{ijk}\vec{\sigma}'_k$ with the totally anti-symmetric Levi-Civita symbol $\epsilon_{ijk}$.
Hence, we obtain
\begin{equation}
  \vec{n} = \sum\limits_{i=1}^3h_i(\vartheta',\varphi')\vec{\beta}\times\vec{\sigma}'_i+\sum\limits_{i,j,k=1}^3\frac{\partial s'_i}{\partial\vartheta'}\frac{\partial s'_j}{\partial\varphi'}\epsilon_{ijk}\vec{\sigma}'_k
\end{equation}
with the abbreviation
\begin{equation}
  h_i(\vartheta',\varphi') = \sum\limits_{j=1}^3\left(\frac{\partial s'_j}{\partial\vartheta'}\frac{\partial s'_i}{\partial\varphi'} - \frac{\partial s'_j}{\partial\varphi'}\frac{\partial s'_i}{\partial\vartheta'}\right)\vec{\sigma}'_j \cdot \left(-\frac{\gamma}{\omega_{\sph}}\vec{\mu}+\frac{\gamma^2}{\gamma+1}\vec{\beta}\right)
\end{equation}
By means of the normal vector, we can construct the equation for the tangent plane $\vec{n}\cdot(\vec{x}-\vec{y})=0$, where the apparent point $\vec{x}$ is the reference point of the plane.
The arbitrary positional vector $\vec{y}$ has to be replaced by the observer position $\vec{x}_{\obs}$. 
The resulting implicit equation for $\vartheta'$ and $\varphi'$ defines the silhouette of the photo-object which has to lie on a right circular cone with apex at the observer.

% -----------------------------------------------------------------
\subsection{Boas strategy}
Another possibility to prove the circular silhouette of a moving sphere starts from within the moving frame $S'$ where the sphere is at rest, see also Boas~\cite{Boas:1961:LORS}.
For that, we first have to transform the observer via the inverse Poincar\'e transformation from $S$ into $S'$, $x'^{\mu}_{\obs} = ?{\bar\Lambda}^{\mu}_{\nu}?\left(x_{\obs}^{\nu}-a^{\nu}\right)$.
\begin{figure}[ht]
  \centering
  \includegraphics[width=0.65\textwidth]{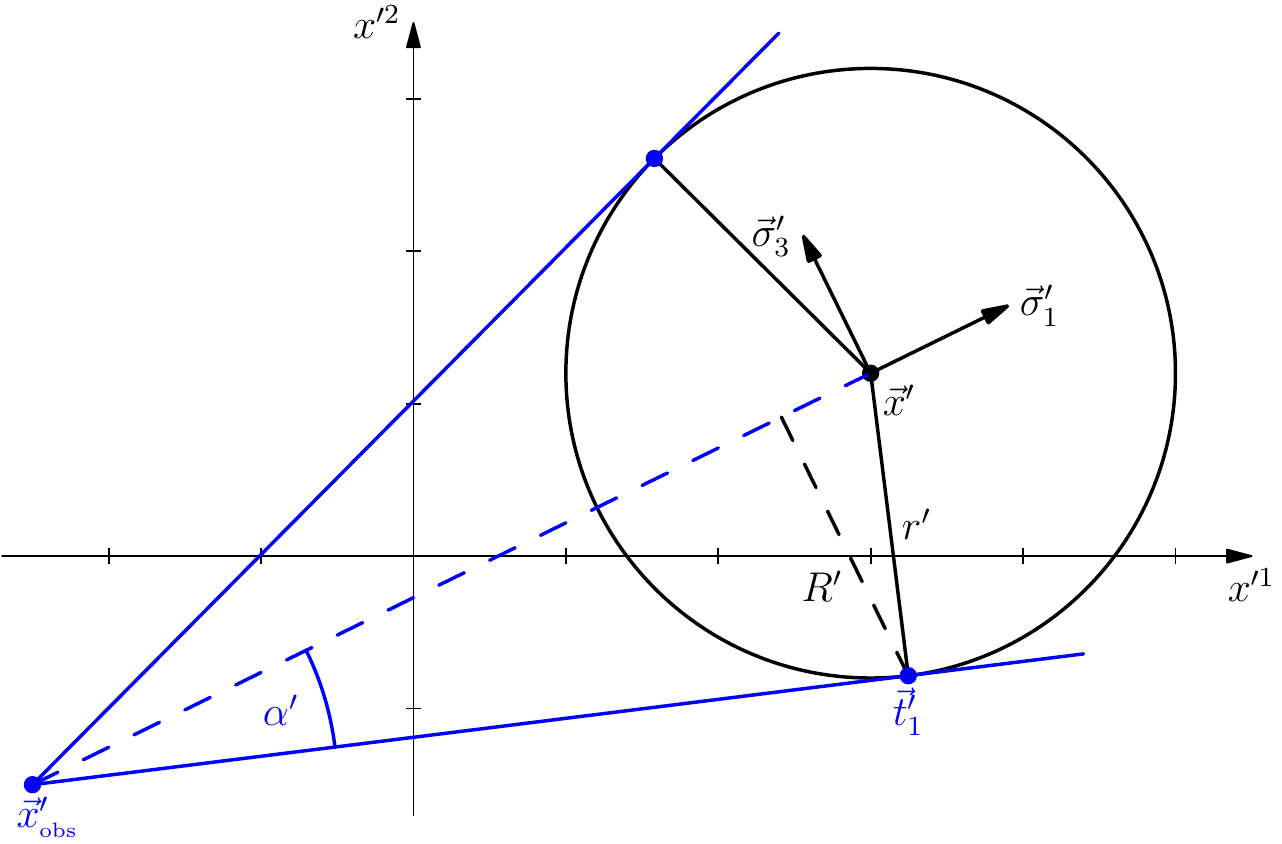}
  \caption{Tangent cone (blue lines) with apex angle $\alpha'$ of a sphere with radius $r'$ and centre point $\vec{x}'$.
  The basis vector $\vec{\sigma}'_2$ points into the plane of projection.
  }
  \label{fig:tangentCone}
\end{figure}

In $S'$, the parameters of the cone tangential to the sphere can be easily determined (see figure~\ref{fig:tangentCone} for a two-dimensional equivalent),
\begin{equation}
  \vec{d}' = \vec{x}' - \vec{x}'_{\obs}, \quad \sin\alpha' = \frac{r'}{\|\vec{d}'\|},\quad R' = r'\cos\alpha',
\end{equation}
where $\vec{d}'$ is the cone axis, $\alpha'$ the apex angle, and $R'$ is the radius of the contact ring.
As the orientation of the sphere has no influence, we can set the basis vectors $\vec{\sigma}'_1$, $\vec{\sigma}'_2$, and $\vec{\sigma}'_3$ as shown in figure~\ref{fig:tangentCone}.
Then, the contact ring $\vec{x}'_{\con}(\psi')=\vec{x}'_{\obs}+\vec{k}'(\psi')$ can be parametrized by the angle $\psi'$ and 
\begin{equation}
  \vec{k}'(\psi') = \frac{r'\cos\alpha'}{\tan\alpha'}\vec{\sigma}'_1 + R'\cos\psi'\vec{\sigma}'_2 + R'\sin\psi'\vec{\sigma}'_3.
\end{equation}
Now, the time $x'^0(\psi')$ when the point $\vec{x}'(\psi')$ has to emit light that reaches the observer at $x'^0_{\obs}$ follows from equation (\ref{eq:intersecTimeS}),
\begin{equation}
  x'^0(\psi') = x'^0_{\obs} - \Delta\left(\vec{x}'(\psi'),\vec{x}'_{\obs}\right) = x'^0_{\obs} - \|\vec{d}'\|\cos\alpha'.
\end{equation}
From that, we can determine the contact ring $\vec{x}_{\con}(\psi')$ with respect to $S$ via the Poincar\'e transformation (\ref{eq:poincareSp2S}),
\begin{equation}
  \fl\vec{x}_{\con}(\psi') = \gamma\left(x'^0_{\obs} - \|\vec{d'}\|\cos\alpha'\right)\vec{\beta} + \vec{x}'_{\obs} + \vec{k}'(\psi') + \frac{\gamma^2}{\gamma+1}\left[\vec{\beta}\cdot\left(\vec{x}'_{\obs}+\vec{k}'(\psi')\right)\right]\vec{\beta} + \vec{a}.
\end{equation}
Again, $\vec{x}_{\con}(\psi')$ has to lie on a right circular cone with apex at the observer.

% -----------------------------------------------------------------
\section{View- and perspective projection}\label{appsec:matrices}
The reference frame $\{\vec{e}_x,\vec{e}_y,\vec{e}_z\}$ of the pinhole camera is defined by the eye point, which corresponds to the observer's position $\vec{x}_{\obs}$, the point-of-interest $\vec{p}$, and a preliminary up-vector $\vec{u}$.
Please note that the pinhole camera looks along the negative $\vec{e}_z$ direction, which is defined by $\vec{e}_z = -(\vec{p}-\vec{x}_{\obs})/\|\vec{p}-\vec{x}_{\obs}\|$. 
The right-axis is given by $\vec{e}_{x}=\vec{u}\times\vec{e}_z/\|\vec{u}\times\vec{e}_z\|$, and finally the corrected up-vector follows from $\vec{e}_y=\vec{e}_z\times\vec{e}_x$.
Thus, the \emph{View} matrix $\mathbbm{V}$, which maps a point into the reference frame of the camera, reads
\begin{equation}
  \mathbbm{V} = \left(\begin{array}{cccc} e_x^1 & e_x^2 & e_x^3 & -x_{\obs}^1\\ e_y^1 & e_y^2 & e_y^3 & -x_{\obs}^2\\ e_z^1 & e_z^2 & e_z^3 & -x_{\obs}^3\\ 0 & 0 & 0 & 1 \end{array}\right).
\end{equation}
The perspective projection emulating the view of a pinhole camera is described by the \emph{Projection} matrix
\begin{equation}
    \mathbbm{P} = \left(\begin{array}{cccc} \frac{1}{a}\cot\frac{\mbox{fov}_y}{2} & 0 & 0 & 0\\ 0 & \cot\frac{\mbox{fov}_y}{2} & 0 & 0\\ 0 & 0 & -\frac{f+n}{f-n} & -2\frac{fn}{f-n}\\ 0 & 0 & -1 & 0 \end{array}\right)
\end{equation}
with aspect ratio $a$, near clipping plane $n$, far clipping plane $f$, and vertical field of view $\mbox{fov}_y$.
Note that, for these matrices, we need homogeneous coordinates $(x,y,z,w)$ and the calculations are done in projective space. 
Then, for our purpose, mapping a point $p=(p^1,p^2,p^3)$ from world space onto the camera's view plane works as follows.
Append the homogeneous coordinate $w=1$ to the point and determine the matrix-matrix-vector multiplication
\begin{equation}
  \mathbf{\hat{p}} = \mathbbm{P}\,\mathbbm{V}\,\mathbf{p}\qquad\mbox{with}\qquad \mathbf{p}=(p^1,p^2,p^3,1)^T
\end{equation}
resulting in the projected point $\mathbf{\hat{p}}$. 
The perspective division $\mathbf{\hat{p}}\mapsto (\hat{p}^1,\hat{p}^2,\hat{p}^3)/\hat{p}^w$ yields the view plane coordinates $v_x=\hat{p}^1/\hat{p}^w$ and $v_y=\hat{p}^2/\hat{p}^w$ with $v_x,v_y\in(-1,1)$. 
The coordinate $v_z=\hat{p}^3/\hat{p}^w$ incorporates depth information of the point.

% -----------------------------------------------------------------
%                           thebibliography
%% -----------------------------------------------------------------
\section*{References}
\bibliographystyle{unsrt}
\bibliography{lit_srvis}

\end{document}